\documentclass[acmsmall,screen,nonacm]{acmart}

\usepackage{graphicx}
\usepackage{caption}
\usepackage{subcaption}
\usepackage{xspace}
\usepackage{adjustbox}
\usepackage{booktabs}
\usepackage{amsmath}
\usepackage{listings}
\usepackage{algorithm}
\usepackage{algorithmicx}
\usepackage{algpseudocode} % This is needed for \algorithmicrequire and \algorithmicensure
\usepackage{multicol}
\usepackage[dvipsnames]{xcolor}
\usepackage{ninecolors}
\usepackage{tikz}
\usepackage{wrapfig}
\usepackage{enumitem}

\newcommand{\myred}{red4}

\newcommand{\system}{SparseConflicts\xspace}
\newcommand{\sparse}[1]{{\textbf{\textit{#1}}}}
\newcommand{\heading}[1]{\vspace{4pt}\textbf{#1}\enspace}

\newcommand{\ir}[1]{\text{\normalfont\texttt{#1}}}
\DeclareMathOperator{\nnz}{nnz}

\algnewcommand{\IIf}[1]{\State\algorithmicif\ #1\ \algorithmicthen}
\algnewcommand{\EndIIf}{\unskip\ \algorithmicend\ \algorithmicif}

\newcommand{\start}{\mathrm{start}}
\newcommand{\curr}{\mathrm{curr}}
\newcommand{\anext}{\mathrm{next}}
\newcommand{\prefix}{\mathrm{prefix}}
\newcommand{\abody}{\mathrm{body}}
\newcommand{\inner}{\mathrm{inner}}

\newcommand{\metacomment}[3]%
{}
\newcommand{\encircle}[1]{%
  \tikz[baseline=(X.base)] 
    \node (X) [draw, shape=circle, inner sep=0] {\strut #1};}

\newcommand{\algscs}{\textsc{Find-All-SCS}\xspace}
\newcommand{\algscslongform}{\textsc{Find All Shortest Common Supersequences}\xspace}
\newcommand{\algcreateaccessmap}{\textsc{Create-Access-Map}\xspace}
\newcommand{\algcreateaccessmaplongform}{\textsc{Create Access Map}\xspace}
\newcommand{\algconvertscstoir}{\textsc{Convert-SCS-To-IR}\xspace}
\newcommand{\algconvertscstoirlongform}{\textsc{Convert SCS to IR}\xspace}

\def\BibTeX{{\rm B\kern-.05em{\sc i\kern-.025em b}\kern-.08em
    T\kern-.1667em\lower.7ex\hbox{E}\kern-.125emX}}

\setcopyright{none}
\acmJournal{PACMPL}
\copyrightyear{2026}
\acmYear{2026}
\acmVolume{X}
\acmNumber{OOPSLA1}
\acmArticle{X}
\acmDOI{x.x}

\begin{document}

% add alias for paper title
\newcommand{\papertitle}{Handling Conflicting Data Layouts in Sparse Tensor Contractions}

\title{\system: \papertitle}
%\subtitle{Generating kernels that do not require explicit transposition of tensors and automatic transformation to nested multi-child/inner loops.}

% \author{Anonymous Author(s)}

%% Author with single affiliation.
\author{Adhitha Dias}
% \authornote{with author1 note}          %% \authornote is optional;
                                        %% can be repeated if necessary
\orcid{0000-0003-3500-7547}             %% \orcid is optional
\affiliation{
  \department{Electrical and Computer Engineering}              %% \department is recommended
  \institution{Purdue University}            %% \institution is required
  \streetaddress{610 Purdue Mall}
  \city{West Lafayette}
  \state{IN}
  \postcode{47906}
  \country{USA}
}
\email{kadhitha@purdue.edu}

\author{Kirshanthan Sundararajah}
\orcid{0000-0001-6384-062X}
\affiliation{
  \department{Computer Science}
  \institution{Virginia Tech}
  \city{Blacksburg}
  \state{VA}
  \country{USA}
  \postcode{24061}
}
\email{kirshanthans@vt.edu}

\author{Artem Pelenitsyn}
\orcid{0000-0001-8334-8106}
\affiliation{
  \department{Electrical and Computer Engineering}
  \institution{Purdue University}
  \streetaddress{610 Purdue Mall}
  \city{West Lafayette}
  \state{IN}
  \country{USA}
  \postcode{47906}
}
\email{apelenit@purdue.edu}

\author{Milind Kulkarni}
\orcid{0000-0001-6827-345X}
\affiliation{
  \department{Electrical and Computer Engineering}
  \institution{Purdue University}
  \streetaddress{610 Purdue Mall}
  \city{West Lafayette}
  \state{IN}
  \country{USA}
  \postcode{47906}
}
\email{milind@purdue.edu}
\begin{abstract}

Optimizing sparse tensor computations is challenging due to the use of compressed storage formats, which leads to non-affine loop nests and a vast, complex schedule space. The performance of a given schedule is sensitive to the sparsity pattern of the input tensors, making it difficult to find a single optimal solution. 
When input tensors in the same tensor contraction have conflicting data layouts in relation to the iteration order, it requires costly---both in time and memory---layout transformation, such as transposition. %\ks{rather than conflicting data layout we can be explicit here. "when the data layouts do not align with the iterations of tensors, it requires costly--- both in performance and memory ---layout transformation(\eg transposition)"}
A promising but under-explored alternative is to generate a schedule that avoids explicit transposition, but this has not been systematically supported in existing compilers.%\ks{Not sure whether fusion is the best term here. I would go with schedule to avoid transposition. Because, fusion makes it sound like a trivial problem.} \ad{"fuse the transposition with the main computation" $\rightarrow$ "is to generate a schedule that avoid explicit transposition."?}
% \ap{transposition was just an example in the previous sentence, but now it's in the center; this is a rough switch.}

This paper presents a new code generation strategy that generalizes the intermediate representation of the TACO sparse tensor compiler to generate a single loop nest, circumventing explicit transposition of tensors when the tensors have conflicting data layout iteration orders. %\ks{Note my comment before about fusion.} 
We extend TACO's iteration graph to express a search-based strategy for locating elements in tensors with conflicting layouts, and we introduce new intermediate representation nodes to lower these schedules to efficient code. This enables the systematic generation of loops that do not require explicit transposition, thus avoiding the overhead of materializing temporary tensors.

We evaluate our approach on a set of sparse tensor contractions using both real-world and synthetic datasets. Our results demonstrate that for computations with misaligned data layouts, our fused approach achieves up to $2\times$ speedup for some sparsity patterns over the traditional approach of explicitly creating a transposed temporary. We also provide guidelines for when this new scheduling strategy is likely to be beneficial.

\end{abstract}

%% 2012 ACM Computing Classification System (CSS) concepts
%% Generate at 'http://dl.acm.org/ccs/ccs.cfm'.
  
%% End of generated code

%% Keywords

\maketitle

\section{INTRODUCTION}\label{spconf:introduction}

Sparse tensor computations are key components in many scientific computing areas, including, but not limited to numerical simulations, data analysis, and machine learning applications~\cite{tce2003, ran2017review, featgraph, fusedmm, tncbook, simquant}.
Sparse tensors are often represented using compressed storage formats such as Compressed Sparse Row (CSR) or Compressed Sparse Fiber (CSF) to save memory and improve computational efficiency~\cite{chou}. 
Memory saving comes from storing only the non-zero elements of the tensor along with their indices, while computation efficiency is achieved by not performing operations on zero elements (i.e., $x \times 0 = 0$ and $x + 0 = x$). 
Using these compressed storage formats makes it challenging to perform tensor operations directly on the compressed data because it requires iteration over the non-zero elements and their indices, which can be complex compared to dense tensor operations~\cite{taco, chou}. 

A key challenge in optimizing sparse tensor computations is minimizing the overhead associated with accessing and manipulating compressed data structures. The use of compressed storage formats in sparse tensor computations leads to \textit{non-affine} loops, where loop bounds and memory accesses depend on metadata arrays. 
These non-affine loop nests prevent the direct application of classical affine loop nest transformation frameworks. 
Consequently, determining an optimal execution schedule\footnote{A schedule is a concrete realization of the computation, which encompasses, for instance, a loop structure, parallelization strategies, tiling/blocking, loop fusion, etc.} for sparse computations is significantly more challenging than for their dense counterparts that operate on affine loops.
This problem has been extensively studied in the compiler literature~\cite{taco, aart22, aartbik:93, ye2023sparsetir, compiler_in_mlir, chou, galley, amir2024-sparse-differentiation, sparseconv2023}, with proposed solutions involving techniques such as fusion, tiling, and data layout transformations, etc~\cite{sparselnr, finch2025, kjolstad:2018:workspaces, senanayake:2020:scheduling, chou2020-format-conversion, mueller:2020:transposition}.

% \ks{This paragraph is not significantly different from the previous one. What is the main idea we are trying to convey here?}\ad{merged with previous paragraph. I think it's good to have more clarity why dense and sparse are different}

The effectiveness of a schedule for non-affine loops is highly sensitive to the sparsity pattern of the input tensors~\cite{waco2023}. 
A loop structure that performs well for one pattern may be inefficient for another. 
Since it is impractical for programmers to manually explore the vast space of possible implementations for every use case, modern sparse tensor compilers provide frameworks that use \textit{domain-specific languages (DSLs) to express a wide range of schedules} without exposing the complexity of low-level details. 
This allows the user to select the most performant schedule for a specific problem instance~\cite{taco, senanayake:2020:scheduling}. %\ks{With the above paragraph, we are moving into the territory of auto scheduling. This is not necessary to our story.} \ad{The majority of the paragraph focuses on saying that no single schedule is optimal for all sparsity patterns, this highlights the need for generating schedules in different ways. Only the last sentence is about auto-scheduling, that is also in combination with "user and auto-scheduler", not directly about auto-scheduling. I italicized the "wide range of schedules" to highlight the main point. I can remove the auto-scheduling part if you think it is not necessary.}

A common scheduling challenge arises when a computation involves tensors with conflicting iteration orders. 
For example, consider an element-wise matrix product\footnote{We use bold letters, like $\sparse A$ and $\sparse B$, to denote sparse matrices or tensors. Dense tensors are not bolded.}: \[{\sparse A(i,j)} \times {\sparse B(i,j)},\] where tensor $\sparse A$ is stored in a row-major format (CSR) and tensor $\sparse B$ is in a column-major format (CSC). Co-iterating over both tensors is difficult due to their conflicting layouts. Specifically, accessing elements of $\sparse A$ requires iterating over index $i$ before $j$, while accessing elements of $\sparse B$ requires iterating over index $j$ before $i$. This means that $\sparse A$ requires a loop order of $i$ outer and $j$ inner, while $\sparse B$ requires $j$ outer and $i$ inner. This mismatch in iteration order is our core problem and complicates the generation of loop code.
% \ks{We need to be explicit about the problem here: mention that $\sparse A$ requires a loop order of $i$ after $j$ while $\sparse B$ requires $j$ after $i$. This is our core problem, We should directly explain it.}
The standard solution is to create a temporary sparse tensor by explicitly transposing one of the inputs to match the other's layout. In this case, converting $\sparse B$ to a row-major format (CSR) to match $\sparse A$'s layout. 
% \ks{Again, be explicit about the solution with the example.}
However, this materialization of a temporary copy incurs time and memory costs, as does the use of alternative structures like hash maps~\cite{sparselnr, const2024}.

\begin{wrapfigure}{r}{0.5\textwidth}
    \includegraphics[width=0.48\textwidth]{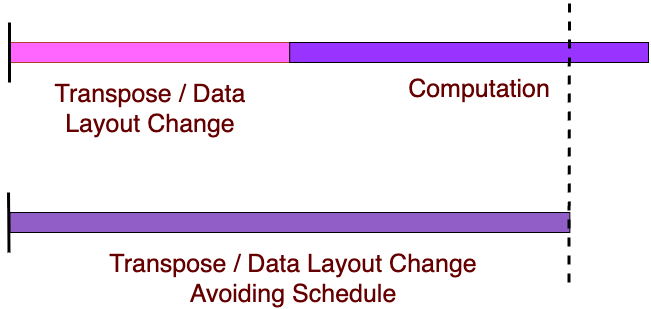}
    \caption{The effect of omitting an explicit transpose step on execution time.}
    \label{spconf:fig:transpose_and_compute}
\end{wrapfigure}

An alternative to materializing a temporary is to generate a schedule that can handle the conflicting iteration orders directly within the computation.
%\ks{I don't think this sends the right message. Fusion sounds like applying SparseLNR over the computation written with transposition.}\ad{I removed the word "fuse" here.}
This can be achieved with a search-based strategy that locates corresponding elements in the compressed data structures on the fly. For example, using a search and a conditional to match corresponding $i, j$ coordinates in $\sparse A$ with $\sparse B$, or vice versa.
This approach eliminates the need for a temporary copy of the entire tensor and may use lower-dimensional dense temporaries to store intermediate results.

By eliminating an explicit transposition, we can reduce the overhead of data manipulation, leading to improved performance. 
This is particularly important for sparse tensors, where the cost of creating temporary tensors can be significant compared to the actual computation. 
Figure~\ref{spconf:fig:transpose_and_compute} illustrates the benefits of a single, fused computation over a separate transpose-then-compute approach.

This search-based mechanism has not been explored systematically as a generic loop generation strategy to eliminate explicit transpositions in sparse tensor computations. %\ks{citation for the ad-hoc manner solutions.} \ad{I changed the wording a bit.}
% While the search-based approach is a known optimization~\cite{intel-mkl-sparseblas}\ks{citation if it is known}, their application to systematically eliminate explicit transpositions in sparse tensor computations has been limited and done in an ad-hoc, schedule specific manner. 
Existing sparse tensor compilers, such as TACO~\cite{taco}, lack a formal mechanism to implement this strategy because their internal representation is over-constrained and not expressive enough to describe such schedules. %\ks{I would say the IR is over constrained. That works well with relaxed iteration graph} 
In this work, we introduce a more generic representation that extends TACO's iteration graph to support a wider range of schedules, including those that avoid explicit transposition. 
Our key idea is to introduce a relaxed iteration graph that can express a search-based strategy, which we then lower to a new intermediate representation for code generation. %\ks{Let's be explicit here. It is an IR more relaxed than the TACO IR. Also, we are using the term "search-based strategy" as something widely known. I don't think it is a familiar term with even for sparse tensor compiler folks.}
This allows us to systematically generate fused transposition schedules for sparse tensor computations.

The contributions of this paper are as follows:

\begin{description}
\item[Relaxed Iteration Graph and IR] We generalize the iteration graph of TACO to support conflicting iteration orders to circumvent explicit transposition steps in the computation and enable more flexible scheduling. To this end, we generalize the intermediate representation to a relaxed version, by introducing two new IR nodes, \texttt{forsome} and \texttt{forsame}.
\item[Novel Code Generation Strategy] We add support for the new IR nodes in the code generation phase, and adapt the existing code generation capabilities of TACO. We further show how the new IR can be transformed into existing IR constructs in certain scenarios. %\ks{The new IR part should be with the previous contribution.}
\item[Demonstrating Performance for Suitable Tensors] We identify two classes of problems where this new code generation strategy can provide substantial speedups.
\end{description}

The rest of the paper is organized as follows. 
We provide the necessary background in Section~\ref{spconf:background}.
We motivate the problem in Section~\ref{spconf:overview}.
The new compiler design is explained in detail in Section~\ref{spconf:detailed_design}.
We discuss implementation details in Section~\ref{spconf:implementation}.
Evaluation of our compiler is presented in Section~\ref{spconf:evaluation}.
We conclude the paper in Section~\ref{spconf:conclusion} with a discussion.

\section{BACKGROUND}\label{spconf:background}

This section provides essential background on sparse tensor access constraints, tensor index notation, iteration graph representation, and scheduling primitives required to understand the rest of the paper.

\subsection{Sparse Tensor Access Constraints}\label{spconf:sparse_tensors}

Sparse tensors can be stored using various compressed data formats, including Compressed Sparse Row (CSR), Sorted Coordinate (Sorted COO), and Compressed Sparse Fiber (CSF), among others.
These formats are abstracted through the {\em level format}, a tree structure that defines the order in which index arrays must be traversed to access an element.
The compressed data formats impose {\em sparse tensor access constraints} based on the required traversal order of their index arrays.
For instance, when ${\sparse A_{ij}}$ is stored in CSR format, the row index must be traversed before accessing the column index, creating a dependency between indices $i$ and $j$ (which traverse the rows and columns of ${\sparse A}$, respectively).
Consequently, the loops containing $i$ and $j$ cannot be arbitrarily reordered.
The TACO Format Abstraction~\cite{chou} provides a detailed description of level formats.

\subsection{Tensor Index Notation for Tensor Contractions}\label{spconf:tensor_contraction}

Tensor contraction operations are described using notation based on the Einstein Summation (Einsum) convention. 
This convention implicitly denotes summation over repeated indices. 
For instance, the expression $A(i,k) = B(i,j)\cdot C(j,k)$ indicates summation over the repeated index $j$ and is equivalent to the standard mathematical notation $A_{ik} = \sum\nolimits_{j} B_{ij} C_{jk}$\footnote{This is the matrix-matrix multiplication operation.}. 
Throughout this text, we use both notations interchangeably. 
This computation can be implemented using a triply nested loop with time complexity $O(IJK)$, assuming dense matrices, where $I$, $J$, and $K$ represent the loop bounds. 
When ${\sparse B}$ is sparse, the time complexity reduces to $O(\nnz({\sparse B_{ij}})K)$, where $\nnz$ denotes the number of non-zero elements.

\subsection{Iteration Graph}\label{spconf:iteration_graph}

% include spmm_iteration_graph.pdf
\begin{figure*}[t]
    % \centering
    % \hspace{-6em}
    \begin{subfigure}{0.25\linewidth}
        \centering
        \includegraphics[width=0.5\linewidth]{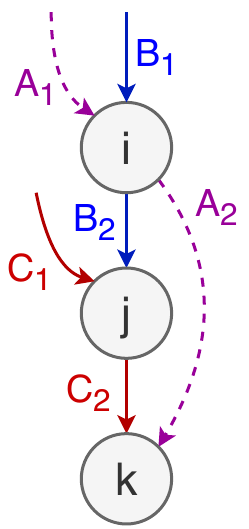}
        % \vspace{1.5em}
        \caption{}
        \label{spconf:fig:spmm_iteration_graph}
    \end{subfigure}
    \hspace{2em}
    \begin{subfigure}{0.60\linewidth}
        \begin{lstlisting}[language=C++,tabsize=2]
for (int32_t i = 0; i < B1_dimension; i++) {
	for (int32_t jB = B2_pos[i]; jB < B2_pos[(i + 1)]; jB++) {
    int32_t j = B2_crd[jB];
    for (int32_t k = 0; k < C2_dimension; k++) {
      int32_t kA = i * A2_dimension + k;
      int32_t kC = j * C2_dimension + k;
      A_vals[kA] = A_vals[kA] + B_vals[jB] * C_vals[kC];
    }
  }
}
        \end{lstlisting}
        % \vspace{-1.5em}
        \caption{}
        \label{spconf:fig:spmm_code}
    \end{subfigure}
    % \vspace{-0.5em}
    \caption{An example of an iteration graph for sparse matrix-matrix multiplication and corresponding code.}
    \label{spconf:fig:spmm-graph-and-code}
\end{figure*}
 
Consider the sparse matrix-matrix multiplication (SpMM) example: $A_{ik} = \sum_{j}\>{\sparse B_{ij}}\>C_{jk}$. 
An iterator traversing indices $i$, $j$, and $k$ reads values from ${\sparse B_{ij}}$ and $C_{jk}$, multiplies corresponding values that share the same $j$ index, and accumulates the results in $A_{ik}$. 
Figure~\ref{spconf:fig:spmm_iteration_graph} shows an example iteration graph, whose internal representation (IR) generates the code shown in Figure~\ref{spconf:fig:spmm_code}. 
In this iteration graph, nodes represent indices from the Einsum notation. 
The graph is acyclic, with edges representing tensor dimensions and their mapping to indices. 
Since ${\sparse B}$ is sparse, the edges $B_{1}$ and $B_{2}$ incident on indices $i$ and $j$ must maintain their order, while other indices can appear in any order when traversing dense tensors (e.g., $C_{1}$ and $C_{2}$). 
No additional sparse tensor access constraints (\S~\ref{spconf:sparse_tensors}) apply in this case. 
TACO provides a detailed description of iteration graphs.

\subsection{Tensor Transposition}\label{spconf:tensor_transpose}

Tensor transposition involves permuting the dimensions of a tensor. For instance, transposing a matrix ${\sparse A_{i,j}}$ produces a new matrix ${\sparse A'_{j,i}}$ with swapped rows and columns.
Sparse tensor transposition is significantly more complex than dense tensor transposition due to the compressed storage formats employed for sparse tensors. In certain cases, transposing a sparse tensor is equivalent to changing its storage format. For the matrix example above, transposition is analogous to converting between CSR and CSC formats. However, in other scenarios, sparse tensor transposition requires reordering the non-zero elements and their corresponding indices within the compressed storage format. This operation can be computationally expensive. Research has explored efficient transposition algorithms using various sorting techniques~\cite{mueller:2020:transposition}, as well as automatic code generation approaches for sparse format conversion~\cite{chou2020-format-conversion}.

\section{OVERVIEW}\label{spconf:overview}

Existing sparse tensor compilers impose specific index access patterns on tensor operands to enable efficient computation. When the required access pattern differs from the tensor's storage format, indices must be reordered through tensor transposition before computation can proceed. While conceptually straightforward, this approach introduces significant memory and computational overheads that can severely impact performance, particularly for large-scale sparse computations.

\subsection{Motivating Example}\label{spconf:motivating_example}

\subsubsection{Element-wise Product with Transpose}

Consider the tensor algebra operation:
\[{\sparse A(i,j)} = {\sparse B(i,j)} * {\sparse C(j,i)}\]

This element-wise multiplication (Hadamard product) computes each element of $\sparse A$ as the product of corresponding elements from $\sparse B$ and $\sparse C$. Such operations are fundamental in tensor factorizations and neural network computations~\cite{algorithm1000, combblas2011}. We assume all tensors are stored in Compressed Sparse Row (CSR) format.\footnote{This computation is equivalent to ${\sparse A(i,j)} = {\sparse B(i,j)} * {\sparse C(i,j)}$ where $\sparse C$ is stored in Compressed Sparse Column (CSC) format while the other tensors remain in CSR format.}

% \ap{See if we can some find references where this kernel is used. or where fusing transpose with compute is useful. If we don't know such an argument, just say one tensor is in csr format and the other tensor is in csc format. }
% \ad{I've mentioned that one tensor can be in CSR and the other one is in CSC format. This is a feasible scenario, and I don't think we need to add more justification.}
% \ad{The other papers don't give exact references and I cannot seem to find a concrete example. So I think this is fine. Commenting this out for now.}

The key challenge arises from mismatched index access patterns: $\sparse B$ requires access order $\encircle{i}\rightarrow\encircle{j}$, while $\sparse C$ follows $\encircle{j}\rightarrow\encircle{i}$. This incompatibility prevents direct computation using existing sparse tensor compilers, which enforce strict access pattern constraints. The conventional solution involves transposing tensor $\sparse C$ to align both operands' access patterns:
\[{\sparse C'(i,j)} = {\sparse C(j,i)}\]
\[{\sparse A(i,j)} = {\sparse B(i,j)} * {\sparse C'(i,j)}\]

While this approach yields the correct result, it introduces significant performance penalties. First, it demands additional memory allocation for the transposed tensor $\sparse C'$, which scales with tensor size and can be prohibitive in memory-constrained environments. Second, it incurs substantial computational overhead for the transpose operation itself, particularly for large sparse tensors. In practice, this approach requires converting the storage format of matrix $C$ from CSR to CSC, followed by the element-wise computation shown in Code Snippet~\ref{spconf:code:hadamard-kernel-taco}.

\begin{figure*}[t]
  %\centering
  \begin{adjustbox}{minipage=\linewidth,scale=0.9}
    \begin{minipage}[t]{0.41\textwidth}
    \centering
    \begin{subfigure}[ct]{0.90\textwidth}
\begin{lstlisting}[basicstyle=\ttfamily\small\linespread{0.5}, tabsize=2, showtabs=false, showstringspaces=false]
iterator i (over A, B, C):
  iterator j (over A, B, C):
    A(i,j) = B(i,j) * C(i,j)
\end{lstlisting}
\caption[]{$ {\sparse A_{ij}} = \sum\nolimits_{} {\sparse B_{ij}} {\sparse C_{ij}} $}
\label{spconf:code:hadamard-kernel-taco}
    \end{subfigure}%
  \end{minipage}%
    \hfill%
    \begin{minipage}[t]{0.55\textwidth}
    \begin{subfigure}[ct]{0.90\textwidth}
\begin{lstlisting}[basicstyle=\ttfamily\small\linespread{0.5}, tabsize=2, showstringspaces=false]
iterator iB (over A, B):
  iterator j (over A, B, C):
    iterator iC (over C):
      if iB == iC:
        A(i,j) = B(i,j) * C(j,i)
\end{lstlisting}
\caption[]{Computation of $ {\sparse A_{ij}} = \sum\nolimits_{} {\sparse B_{ij}} {\sparse C_{ji}} $ without transposing $ {\sparse C_{ji}} $.}
\label{spconf:code:hadamard-kernel-fused}
\Description[]{}
    \end{subfigure}%
  \end{minipage}
  \centering
  \end{adjustbox}
    \caption{(a) Computing $A(i,j)=B(i,j)*C(i,j)$. $A(i,j)=B(i,j)*C(j,i)$ cannot be computed since $B$, and $C$ have conflicting index access patterns. (b) shows how we can transpose from CSR format to CSC format to have the $i, j$ indexing pattern. (c) shows the $A(i,j)=B(i,j)*C(j,i)$ computation.}
    \vspace{-10pt}
    \label{spconf:fig:three graphs}
\end{figure*}

However, we can avoid the transpose operation overheads by directly computing the Hadamard product through a search operation over the index $i$ of $\sparse C$ to match the index $i$ of $\sparse B$. Code Snippet~\ref{spconf:code:hadamard-kernel-fused} demonstrates this approach. While iterating over the same index $i$ of both $\sparse B$ and $\sparse C$ introduces additional computational overhead, the iteration over $\sparse C$ traverses only non-zero element indices. Moreover, CSR format maintains sorted indices within each row, enabling binary search to locate matching index $i$ in $\sparse C$ for each index $i$ in $\sparse B$. This optimization reduces the time complexity of each index search operation from $O(n)$ to $O(\log n)$, where $n$ represents the average number of non-zero elements in a row of $\sparse C$. The relative performance between these scheduling approaches depends on the sparsity attributes and patterns of the tensors, as we will demonstrate in Section~\ref{spconf:evaluation}. 

In these two cases, the transpose-plus-compute approach has a time complexity of $O(\nnz({\sparse C}) + \nnz({\sparse B} \cup {\sparse C}))$, which includes the cost of transposing $\sparse C$ and performing the element-wise multiplication. Here, we use $\nnz({\sparse B} \cup {\sparse C})$ to denote the union of the non-zero elements in $\sparse B$ and $\sparse C$, because the iterators traverse the union of the indices of non-zero elements in both tensors. In contrast, the fused approach has a time complexity of $O(\nnz({\sparse B}) * \log ({\sparse C}_{row-avg}))$, where ${\sparse C}_{row-avg}$ is the average number of non-zero elements per row in $\sparse C$. The fused approach can be more efficient when $\sparse C$ has a high sparsity level, making the transpose operation relatively expensive.

\subsubsection{Sparse Matrix-Sparse Matrix Multiplication}

Consider the tensor algebra operation:
\[{\sparse A(i,j)} = {\sparse B(i,k)} * {\sparse C(k,j)}\]

where $\sparse A$, $\sparse B$, and $\sparse C$ are stored in CSR format. This operation represents sparse matrix-matrix multiplication and is widely used in various applications~\cite{spgemm-hypersparse, combblas2011}. 

When the output of a tensor contraction is sparse and stored in a level-format, the output tensor indices must be positioned as the outermost indices in the compute loop nest. For example, in the above contraction, when indices $i$ and $j$ are the outermost indices in that order, the $(i,j)$ value of the output tensor can be computed by contracting over the $k$ index of the input tensors. If no values exist to contract, the $(i,j)$ value of the output tensor is zero and can be omitted from the output tensor in both the indexing arrays and the value array. Code snippet~\ref{spconf:code:spmm-unoptimized} outlines the basic loop structure of this computation. Similar to the previous example, these loop nests do not contradict the index access patterns of the input tensors or the output tensor. Although the iteration order of $i$ followed by $k$ followed by $j$ does not contradict the access constraints of each individual tensor, we cannot generate this loop structure because the output tensor $\sparse A$ is in CSR format. It requires the outermost indices to be $i$ followed by $j$ so that we can extend the indexing arrays of $\sparse A$ in the correct order.

% \ap{make the point that working in the lower level representation is harder.}

\begin{figure*}[t]
%\centering
\begin{adjustbox}{minipage=\linewidth,scale=0.9}
\begin{minipage}[t]{0.45\textwidth}
\centering
\begin{subfigure}[t]{0.99\textwidth}
\begin{lstlisting}[basicstyle=\ttfamily\small\linespread{0.5}, tabsize=2, showtabs=false, showstringspaces=false]
iterator i (over A, B):
  iterator jA (over A):
    float sum = 0;
    bool found = false;
    iterator k (over B, C):
      iterator jC (over C):
        if jA == jC:
          found = true;
          sum += B(i,k)*C(k,jC);
    if found:
    // extend A's indexing arrays
    A(i,jA) = sum;
\end{lstlisting}
\caption[]{\system pseudocode for ${\sparse A_{ij}} = \sum\nolimits_{} {\sparse B_{ij}} C_{ij} $}
\label{spconf:code:spmm-unoptimized}
\end{subfigure}%
\end{minipage}%
\hfill%
\begin{minipage}[t]{0.45\textwidth}
\begin{subfigure}[t]{0.99\textwidth}
% \begin{lstlisting}[basicstyle=\ttfamily\small\linespread{0.5}, lineskip=1pt, tabsize=2, columns=fullflexible, showstringspaces=false]
% // check me
% float* vals = (float*)malloc(J);
% int* indices = (int*)malloc(J);
% bool* ind_set = (bool*)calloc(J);
% loop i (over A, B):
%   int num_indices = 0;
%   loop k (over B, C):
%     loop j (over C):
%       if !ind_set[j]:
%         ind_set[j] = true;
%         indices[num_indices] = j;
%         vals[j] = B(i,k)*C(k,j);
%         num_indices++;
%       else:
%         vals[j] += B(i,k)*C(k,j);
%   sort(indices, num_indices);
%   loop t (over `indices` to form A): 
%     A(i,indices[t]) = vals[indices[t]];
%     ind_set[indices[t]] = false; 
% \end{lstlisting}
\begin{lstlisting}[basicstyle=\ttfamily\small\linespread{0.5}, lineskip=1pt, tabsize=2, columns=fullflexible, showstringspaces=false]
set<int> relevant_js;
iterator i (over A, B):
  float vals[J];
  iterator k (over B, C):
    iterator j (over C):
      relevant_js.insert(j);
      vals[j] += B(i,k)*C(k,j);
  iterator j (over A): // only relevant_js
    A(i,j) = vals[j];
    relevant_js.remove(j);
\end{lstlisting}
\caption[]{Transformed pseudocode to a nested multi-child/inner loop structure.}
\label{spconf:code:spmm-optimized}
\Description[]{}
\end{subfigure}%
\end{minipage}
\end{adjustbox}
\caption{Computing $A(i,j)=B(i,k)*C(k,j)$ using \system IR and transforming it to an optimized version.}
\label{spconf:fig:spmm-example}
\end{figure*}

Although this approach is functionally correct, it is highly inefficient. The inefficiency stems from the need to iterate over all possible $j$ indices to populate the output tensor's indexing arrays in the correct order. This requirement leads to two separate iterations over the $j$ index, with one of them traversing all possible indices, resulting in a sub-optimal time complexity of $O(I \cdot J \cdot \nnz(C))$. This computation can be optimized by restructuring the loop nest and introducing a temporary workspace. The optimized code, shown in Code Snippet~\ref{spconf:code:spmm-optimized}, computes all values for a single row of the output tensor (i.e., for a single instance of $i$) within the inner $k$ and $j$ loops. The temporary array accumulates the partial results corresponding to the non-zero columns in $\sparse C$ for the current output row $i$.

While general-purpose sparse compilers like TACO~\cite{taco} support computation mechanisms similar to the one shown in Code Snippet~\ref{spconf:code:spmm-optimized}, they require users to define computations at a lower level of abstraction. Consequently, users cannot rely solely on high-level Einsum notation and must instead work with a more complex lower-level IR. To circumvent this complexity for common operations, TACO's implementation includes a hardcoded SpMM kernel that generates the optimized code from Code Snippet~\ref{spconf:code:spmm-optimized} using low-level primitives.

\subsection{Our approach: \system}
\label{spconf:sec:sparsetda}

The insights drawn from the motivating examples and our approach to code generation can be summarized as follows.

\heading{Constraints on Initial Index Layout}
We use sparse access constraints to build the initial layout of the loop iterators (\S~\ref{spconf:relaxed_iteration_graph}).

\heading{A Leader-Follower Model}
We use the initial index layout to identify leader and follower nodes in the iteration graph and build a relaxed iteration graph representation that allows more flexible index traversal orders (\S~\ref{spconf:relaxed_iteration_graph}).
This relaxed iteration graph serves as the foundation for generating a new intermediate representation (IR) that supports more adaptable code generation strategies (\S~\ref{spconf:internal_representation}).

\heading{Transformation Rules for New IR}
We introduce transformation rules to convert the new IR into existing IR constructs when specific conditions are met, enabling the reuse of established code generation techniques (\S~\ref{spconf:transformations_to_ir}).

\section{DESIGN OF THE TRANSFORMATION}\label{spconf:detailed_design}

To support tensor contractions where operand storage formats are incompatible, we introduce a new transformation algorithm. Our approach generates an efficient loop nest by treating the index order as a modified Shortest Common Supersequence (SCS) problem~\cite{shortestcommonsupersequence}. The algorithm proceeds in three phases:

\begin{enumerate}
    \item \textbf{Iteration Graph Scheduling:} We first find the Shortest Common Supersequence (SCS) of the index orders of all tensors in the expression (\S~\ref{spconf:relaxed_iteration_graph}). This SCS defines the loop nest order. For sparse output tensors, we constrain the SCS to begin with the output tensor's free indices\footnote{Free indices are those that appear in the output tensor.} to enable in-order generation of the output. For dense outputs, no such constraint is imposed, allowing for more flexible loop orderings.
    \item \textbf{IR Generation:} The scheduled SCS is then lowered to a new Internal Representation (IR) that extends TACO's concrete index notation (\S~\ref{spconf:internal_representation}). This IR introduces new iterators to explicitly represent iteration over subsets of tensors and to model searches required when an index appears in multiple positions in the loop nest.
    \item \textbf{IR Transformation:} Finally, we apply transformations to the new IR to optimize the loop structure (\S~\ref{spconf:transformations_to_ir}). For instance, we can transform a linear loop nest into a branched structure with producer-consumer semantics. This is particularly effective for sparse outputs where an index appears only in the output tensor, avoiding unnecessary computation.
\end{enumerate}
We now describe each phase in detail.

\begin{figure*}[t]
    \centering
    \begin{subfigure}{0.50\textwidth}
        \centering
        \includegraphics[width=0.90\linewidth]{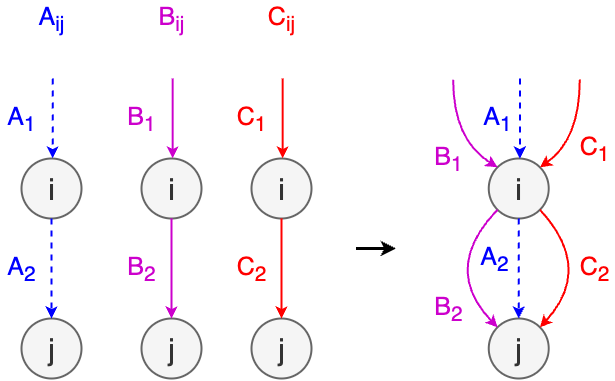}
        \vspace{2em}
        \caption{${\sparse A(i,j)} = {\sparse B(i,j)} * {\sparse C(i,j)}$ default iteration graph.}
        \label{spconf:fig:hadamard-default}
    \end{subfigure}%
    \hfill
    \begin{subfigure}{0.47\textwidth}
        \centering
\begin{lstlisting}[basicstyle=\footnotesize\linespread{0.1}, tabsize=2, columns=fullflexible, showstringspaces=false, escapechar=|]
for (int i = 0; i < B1_dimension; i++) { |\label{spconf:line-i-start}|
  int jB = B2_pos[i]; int pB2_end = B2_pos[i+1];
  int jC = C1_pos[i]; int pC2_end = C2_pos[i+1]; |\label{spconf:line-i-end}|
  while (jB < pB2_end && jC < pC2_end) { |\label{spconf:line-j-loop-start}|
    int jB0 = B2_crd[jB]; int jC0 = C2_crd[jC];
    int j = min(jB0, jC0);
    if (jB0 == j && jC0 == j) { |\label{spconf:line-j-match}|
      A_vals[jA] += B_vals[jB] * C_vals[jC]; jA++; }
    jB += (int) (jB0 == j); |\label{spconf:line-jB-increment}|
    jC += (int) (jC0 == j); |\label{spconf:line-jC-increment}|
  } 
}
\end{lstlisting}
        \caption{Generated code from the graph in Figure~\ref{spconf:fig:hadamard-default}.}
        \label{spconf:fig:hadamard-default-code}
    \end{subfigure}
    \vskip\baselineskip
    \begin{subfigure}{0.50\textwidth}
      \centering
      \includegraphics[width=0.95\linewidth]{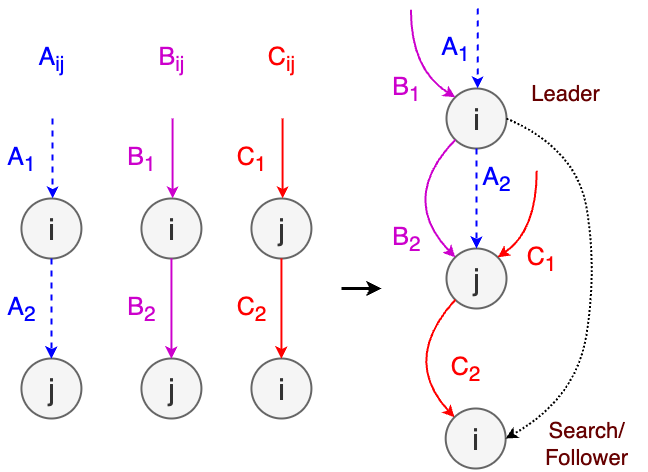}
      \caption{${\sparse A(i,j)} = {\sparse B(i,j)} * {\sparse C(j,i)}$ relaxed iteration graph.}
      \label{spconf:fig:hadamard-fused}
    \end{subfigure}
    \hfill
    \begin{subfigure}{0.47\textwidth}
        \centering
\begin{lstlisting}[basicstyle=\footnotesize\linespread{0.1}, tabsize=2, columns=fullflexible, showstringspaces=false, escapechar=|]
int jA = 0;
for (int i = 0; i < B1_dimension; i++) { |\label{spconf:line-i-start-2}|
  for (int jB = B2_pos[i]; jB < B2_pos[i+1]; jB++) { |\label{spconf:line-jB-loop-start-2}|
    int j = B2_crd[jB]; |\label{spconf:line-j-loop-start-2}|
    int iC = binary_search_index(|\hspace{1em}|idx_array=C2_crd, start=C2_pos[j], |\hspace{1em}|end=C2_pos[j+1], match=i); |\label{spconf:line-iC-search-2}|
    if (iC != -1) { |\label{spconf:line-iC-check-2}|
      A_vals[jA] += B_vals[jB] * C_vals[iC]; 
      jA++;
    }
  }
}
\end{lstlisting}
        \caption{Generated code from the graph in Figure~\ref{spconf:fig:hadamard-fused}.}
        \label{spconf:fig:hadamard-fused-code}
    \end{subfigure}
    \caption{Iteration graphs and generated code for the hadamard product of two sparse matrices.}
    \label{spconf:fig:detailed-design-hadamard}
\end{figure*}

\subsection{Relaxed Iteration Graph Representation}\label{spconf:relaxed_iteration_graph}

Let's consider our previous example of the Hadamard product of the two matrices,
\[{\sparse A(i,j)} = {\sparse B(i,j)} * {\sparse C(i,j)}\]
The output is generated by iterating over $\sparse B$ and $\sparse C$, multiplying values that share the same $i$ and $j$ indices, and storing the result in $\sparse A$. In its simplest form, we can think of the iterations over the input and output tensors individually, as shown in Figure~\ref{spconf:fig:hadamard-default}, and then combine them into a single iteration graph. This iteration graph can be converted into code, as shown in Figure~\ref{spconf:fig:hadamard-default-code}, by mapping each index in the graph to loops in the code (Lines~\ref{spconf:line-i-start}--\ref{spconf:line-i-end} correspond to the $i$ index, while Lines~\ref{spconf:line-j-loop-start}--\ref{spconf:line-j-match} and \ref{spconf:line-jB-increment}--\ref{spconf:line-jC-increment} correspond to the $j$ index). The indexing arrays of the input tensors can be co-iterated to traverse their non-zero elements.

% \ap{this paragraph, after showing the updated kernel, should first say that before this work no one showed how to combine individual iterations graphs (or some such) and that we propose to do X and Y to achieve this} 
% \ad{added a sentence to say that no systematic method existed before our work.}
However, if the computation is
\begin{equation}
{\sparse A(i,j)} = {\sparse B(i,j)} * {\sparse C(j,i)},\label{spconf:eqn:hadamard-trans}
\end{equation}
the individual iteration graphs can be combined into a single graph while keeping all sparse iteration constraints intact, as shown in Figure~\ref{spconf:fig:hadamard-fused}. Before our work, no systematic method existed for combining individual iteration graphs in such scenarios. We introduce a novel approach that models the problem as finding a Shortest Common Supersequence (SCS) of the tensor index orders, which defines a unified loop nest. This, in turn, allows us to create a single, combined iteration graph. The subsequent paragraphs describe our approach to this problem, simplifying it by explaining the concepts using this example.

In this example, the duplicated $i$ node becomes a search node, where the iterator must find the corresponding instance of $i$ from the outermost loop. This mechanism can be viewed as a leader-follower approach: the first instance of the index (the \emph{leader}) iterates freely over the tensors it is incident upon, while the second instance (the \emph{follower}) finds the same value by searching the indexing array of the other tensor. The code generated from this iteration graph is shown in Figure~\ref{spconf:fig:hadamard-fused-code}. Note that line~\ref{spconf:line-i-start-2} corresponds to the first $i$ index, lines~\ref{spconf:line-jB-loop-start-2}--\ref{spconf:line-j-loop-start-2} to the $j$ index, and lines~\ref{spconf:line-iC-search-2}--\ref{spconf:line-iC-check-2} to the second $i$ index. Line~\ref{spconf:line-iC-search-2} performs a binary search over the indexing array of $\sparse C$ to find the instance of $i$ from the outermost loop. The conditional line~\ref{spconf:line-iC-check-2} uses the output from the binary search to decide whether to continue the computation. By combining the iteration graphs in this manner, we can avoid an explicit transposition of $\sparse C$ to match the computation patterns supported by existing sparse tensor compilers.

The algorithm to build the resulting iteration graph for~\eqref{spconf:eqn:hadamard-trans} proceeds as follows. The constraints imposed by each tensor in the computation are collected. In particular, both ${\sparse A(i,j)}$ and ${\sparse B(i,j)}$ are sparse and in CSR format, which imposes the constraint that the $j$ index must follow the $i$ index in the loop nest. Similarly, for ${\sparse C(j,i)}$, the $i$ index must follow the $j$ index. Together, these yield two constraints: $\{\encircle{i} \rightarrow \encircle{j}, \encircle{j} \rightarrow \encircle{i}\}$. Conforming to these constraints allows for two possible index orders: $\encircle{i} \rightarrow \encircle{j} \rightarrow \encircle{i}$ and $\encircle{j} \rightarrow \encircle{i} \rightarrow \encircle{j}$. However, because the output tensor is sparse, the order $\encircle{j} \rightarrow \encircle{i} \rightarrow \encircle{j}$ must be eliminated, as it would prevent the output indexing arrays from being generated in the correct order if the indices are not placed as the outermost indices in the loop nest.

As another example, consider the contraction $a(i) = {\sparse B(i,j,k)} * {\sparse C(i,k,j)}$, where the output $a$ is dense, and the inputs $\sparse B$ and $\sparse C$ are in CSF format. The constraints collected from the tensors are $\{\encircle{i} \rightarrow \encircle{j}, \encircle{j} \rightarrow \encircle{k}\}$ from $\sparse B$ and $\{\encircle{i} \rightarrow \encircle{k}, \encircle{k} \rightarrow \encircle{j}\}$ from $\sparse C$. Two possible indexing orders that conform to these constraints are $\encircle{i} \rightarrow \encircle{j} \rightarrow \encircle{k} \rightarrow \encircle{j}$ and $\encircle{i} \rightarrow \encircle{k} \rightarrow \encircle{j} \rightarrow \encircle{k}$. Both are valid because the output is dense, so the free index $i$ does not need to be fixed at the beginning of the index order.

Algorithm~\ref{spconf:alg:scs} describes a procedure to find all possible index orders that conform to the given constraints. We describe the formal model and provide proof sketches for optimality and completeness of the algorithm below.

Notation:
\begin{itemize}
    \item $S = \{S_1, S_2, \ldots, S_n\}$: the set of input sequences where each sequence is a sparse access constraint. %\ap{sequence of what? what kind of input?}. \ad{added "sparse access constraint"}
    \item $\Sigma_{\prefix}$: the required fixed prefix sequence (if any).
    \item $\mathcal{R}$: the set of resulting SCS.
    \item $p = (p_1, p_2, \ldots, p_n)$: a state vector where $p_i$ is the current index in sequence $S_i$.
    \item $p_{term} = (|S_1|, |S_2|, \ldots, |S_n|)$: the terminal state where all sequences are fully consumed.
    \item $\Pi$: the current supersequence being constructed, or a path representing a partially or fully constructed supersequence.
    \item $\mathcal{Q}$: a queue for the BFS, storing start node objects.
    \item $D$: a map storing the length of the shortest path found to a state, i.e., $D[p]$ is the length of the shortest path found to state $p$.
    % \item $|\Pi|$: The length of the sequence $\Pi$.
    \item $\Pi\cdot c$: concatenation of character $c$ with sequence $\Pi$.
\end{itemize}

\algrenewcommand\algorithmicindent{1.0em}% % or whatever size you prefer

\begin{algorithm}
\caption{\algscslongform}
\label{spconf:alg:scs}
\begin{multicols}{2}
\footnotesize{
\begin{algorithmic}[1]
\Procedure{\algscs}{$S, \Sigma_{\prefix}$} returns $\mathcal{R}$
\State $\mathcal{R} \leftarrow \emptyset$
\State $\mathcal{Q} \leftarrow \text{Queue()}$
\State $D \leftarrow \text{Map()}$
\State $L_{\min} \leftarrow \infty$
\State $n \leftarrow |S|$ \Comment{Number of sequences}

\State \Comment{Calculate the initial state based on the prefix}
\State Let $\vec{p}_{\start}$ be a new state vector of size $n$ \label{spconf:line:initialize-p-start}
\For{$i \leftarrow 1$ to $n$}
    \State $p_{\start, i} \leftarrow$ length of the longest prefix of $S_i$ that is a subsequence of $\Sigma_{\prefix}$.
\EndFor \label{spconf:line:initialize-p-end}

\State 
\State $\text{node}_{\start} \leftarrow (\vec{p}_{\start}, \Sigma_{\prefix})$ \label{spconf:line:initialize-bfs-start} \Comment{Initialize the BFS}
\State $\mathcal{Q}.\text{Enqueue}(\text{node}_{\start})$
\State $D[\vec{p}_{\start}] \leftarrow |\Sigma_{\prefix}|$ \label{spconf:line:initialize-bfs-end}

\State \Comment{Main BFS Loop}
\While{$\mathcal{Q}$ is not empty}
    \State $(\vec{p}_{\curr}, \Pi_{\curr}) \leftarrow \mathcal{Q}.\text{Dequeue}()$ \label{spconf:line:dequeue}
    \If{$|\Pi_{\curr}| \ge L_{\min}$} \label{spconf:line:pruning-check} \Comment{Pruning check}
        \State \textbf{continue}
    \EndIf
    \columnbreak % Forces a break to the next column
    \If{$\vec{p}_{\curr} = \vec{p}_{term}$} \label{spconf:line:goal-check} \Comment{Goal check}
        \If{$L_{\min} = \infty$} \label{spconf:line:first-solution-start}
            \State $L_{\min} \leftarrow |\Pi_{\curr}|$
        \EndIf \label{spconf:line:first-solution-end}
        \State Add $\Pi_{\curr}$ to $\mathcal{R}$; \label{spconf:line:add-to-results} \textbf{continue}
    \EndIf

    \State \Comment{Generate successor states}
    \State $C \leftarrow \{S_i[p_{\curr, i}] \mid 1 \le i \le n \text{ and } p_{\curr, i} < |S_i|\}$
    \For{each character $c \in C$} \label{spconf:line:for-each-char-start}
        \State Let $\vec{p}_{\anext}$ be a new state vector
        \For{$i \leftarrow 1$ to $n$}
            \If{$p_{\curr, i} < |S_i|$ and $S_i[p_{\curr, i}] = c$}
                \State $p_{\anext, i} \leftarrow p_{\curr, i} + 1$
            \Else
                \State $p_{\anext, i} \leftarrow p_{\curr, i}$
            \EndIf
        \EndFor
        \State $\Pi_{\anext} \leftarrow \Pi_{\curr} \cdot c$

        \If{$\vec{p}_{\anext} \notin D$ or $|\Pi_{\anext}| \le D[\vec{p}_{\anext}]$} \label{spconf:line:preserve-parallel-paths}
            \State $D[\vec{p}_{\anext}] \leftarrow |\Pi_{\anext}|$
            \State $\mathcal{Q}.\text{Enqueue}((\vec{p}_{\anext}, \Pi_{\anext}))$
        \EndIf
    \EndFor \label{spconf:line:for-each-char-end}
\EndWhile
\State \textbf{return} $\mathcal{R}$
\EndProcedure
\end{algorithmic}
}
\end{multicols}
\end{algorithm}

% \begin{description}[leftmargin=0cm]

\textbf{Specification} We model the problem on a state-space graph $G = (V, E)$. %\ap{I don't think it's a good idea to call this a ``formal model''; it looks like a \emph{Specification} of the Algorithm maybe: it describes its inputs and outputs. Also, there should be a paragraph giving a verbal explanation of the pseudocode (I know you're always reluctant to include it! but it's safer to have it). Also, I'm not a fan of lists with plenty of text and even less so of such nested lists. The text normally should be just text with sections/subsections/paragraps (speaking in latex jargon here).} \ap{something went wrong with punctuation and grammar here} 
% \ad{I had this in a paragraph before, and thought it would be clearer in a list, because there are a lot of annotations/symbols here. It will be more convoluted in paragraph form. I use examples to explain the procedures, and the proof sketches explain bits of the algorithm in more detail.}
% \ad{rephrased to "We model the problem on a state-space graph G = (V, E)". This happened because this was a paragraph before.}
% \ad{removed "formal model" and replaced with "specification"}
% \ad{removed nested list}
\begin{itemize}
    \item A vertex $v \in V$ represents a state vector $\vec{p} = \langle p_1, p_2, \ldots, p_n \rangle$ indicating the current position in each sequence $S_i$.
    \item An edge $(v_{a}, v_{b}) \in E$ with label $c$ exists if state $\vec{p}_{a}$ is reachable from $\vec{p}_{b}$ by appending character $c$ to the supersequence (lines~\ref{spconf:line:for-each-char-start}--\ref{spconf:line:for-each-char-end}), advancing the position in each sequence $S_i$ that has $c$ at its current position.
    \item The problem is to find all shortest paths from a starting vertex $v_{\prefix}$ to the terminal vertex $v_{term}$. The vertex $v_{\prefix}$ corresponds to the state vector $\vec{p}_{\start}$ computed in lines~\ref{spconf:line:initialize-p-start}--\ref{spconf:line:initialize-p-end} of the algorithm.
    \item The final supersequence is the concatenation of $\Sigma_{\prefix}$ and the labels of a path from $v_{\prefix}$ to $v_{term}$. Minimizing the total length is equivalent to minimizing the length of the path from $v_{\prefix}$ to $v_{term}$.
\end{itemize}

% \item % for spacing

% \ks{Let's go with proof sketches}

\textbf{Proof Sketch of Correctness (Optimality)} The algorithm \algscs{} finds a supersequence that both starts with $\Sigma_{\prefix}$ and has the shortest possible total length. The proof of correctness is based on the following observations: %\ap{the name of the algo should be a macro: you already have different names: the listing calls it just SCS} \ad{added a macro. corrected the algorithm title.}

\begin{itemize}
  \item \textit{Objective Equivalence:} Minimizing the total length of the supersequence is equivalent to finding a shortest path in the unweighted graph $G$ from the vertex $v_{\prefix}$ to $v_{term}$
  \item \textit{BFS Implementation:} The algorithm implements a Breadth-First Search (BFS) strategy to explore the state-space graph $G$. The queue $\mathcal{Q}$ is used to explore states level by level, ensuring that the first time a state is reached, it is via the shortest path. It begins the search at the single node $(\vec{p}_{\start}, \Sigma_{\prefix})$ as initialized in lines~\ref{spconf:line:initialize-bfs-start}--\ref{spconf:line:initialize-bfs-end}. % \ap{ranges should use two hyphens in the latex source not one; double-check everywhere!}. \ad{range fixed}
  \item \textit{BFS Optimality Property:} BFS guarantees that the first time it reaches any vertex $v$ from a source, it has found a shortest path from the source to $v$. In our case, this means that when the algorithm first reaches the terminal state $\vec{p}_{term}$, it has found a shortest path from $\vec{p}_{\prefix}$ to $\vec{p}_{term}$.
  \item \textit{First Solution:} The algorithm finds its first solution when it processes a node $(\vec{p}_{\curr}, \Pi_{\curr})$ where $\vec{p}_{\curr} = \vec{p}_{term}$ (line~\ref{spconf:line:goal-check}). At this point, $|\Pi_{\curr}|$ is the length of the shortest path from $\vec{p}_{\prefix}$ to $\vec{p}_{term}$, and $L_{\min}$ is set to this length (lines~\ref{spconf:line:first-solution-start}--\ref{spconf:line:first-solution-end}).
  \item \textit{Conclusion:} The length of the first solution is $|\Sigma_{\prefix}| + L'$, where $L'$ is the length of the shortest path from $\vec{p}_{\prefix}$ to $\vec{p}_{term}$. Since BFS guarantees $L'$ is minimal, the total length stored in $L_{\min}$ is the shortest possible for any valid supersequence that adheres to the prefix constraint.
\end{itemize} 

% \item % for spacing

\textbf{Proof Sketch of Completeness (Finding All Shortest Paths)} The algorithm \algscs{} finds all distinct supersequences of the shortest possible length that begin with $\Sigma_{\prefix}$. The proof of completeness is based on the following observations:

\begin{itemize}
  \item \textit{Assumption for Contradiction:} Assume there exists a valid SCS, $\Sigma^{*}$, of the minimal length $L_{\min}$ which the algorithm fails to add to the result set $\mathcal{R}$.
  \item \textit{Path Correspondence:} $\Sigma^{*}$ corresponds to a specific shortest path of length $L' = L_{\min} - |\Sigma_{\prefix}|$ from the vertex $v_{\prefix}$ to $v_{term}$ in the state-space graph $G$.
  \item \textit{Systematic Exploration:} The BFS systematically explores all possible paths of length $k$ from $v_{\prefix}$ before any of length $k+1$. The pruning condition (line~\ref{spconf:line:pruning-check}) ensures the algorithm only stops extending the paths after they have exceeded the known shortest length, $L_{\min}$, but it fully explores all paths of length $L_{\min}$.
  \item \textit{Path Discovery/Preservation of Parallel Paths:} The crucial condition on line~\ref{spconf:line:preserve-parallel-paths} ensures that if the algorithm discovers a new path to an existing state $\vec{p}_{\anext}$ that is equally short as a previously discovered path, it still enqueues this new path $(\vec{p}_{\anext}, \Pi_{\anext})$ for further exploration. This mechanism preserves all parallel paths of the same length, ensuring that no potential shortest path is overlooked.
  \item \textit{Contradiction and Conclusion:} Because the algorithm explores all equally short paths to any intermediate state, it is guaranteed to discover \textbf{all} shortest paths from $v_{\prefix}$ to $v_{term}$. Each such path corresponds to a unique state node $\vec{p}_\curr$ that will be processed at line~\ref{spconf:line:dequeue}, and its path $\Pi_{\curr}$ will be added to the result set $\mathcal{R}$ (line~\ref{spconf:line:add-to-results}). Therefore, the assumption that a valid shortest solution could be missed is contradicted, proving that the algorithm is complete in finding all shortest supersequences.
\end{itemize}

% \end{description}

While our method selects one of these valid orders, the selection of the optimal index order is left as future work. Furthermore, any indices not constrained by sparse tensor access patterns can be inserted at any position in the chosen order.

\begin{figure*}
\centering
% \begin{multicols}{2}
{\footnotesize
\begin{subfigure}{0.40\textwidth}
\begin{align*}
\ir{Stmt} \quad ::=& \quad \ir{Assignment} \\
|& \quad \ir{Forall} \\
|& \quad \ir{\textcolor{\myred}{Forsome}} \\
|& \quad \ir{\textcolor{\myred}{Forsame}} \\
|& \quad \ir{Where} \\
\ir{Assignment} \quad ::=& \quad \ir{Access} = \ir{Expr} \\
|& \quad \ir{Access} \text{ += } \ir{Expr} \dots
\end{align*}
\end{subfigure}
\begin{subfigure}{0.40\textwidth}
\begin{align*}
\ir{Forall} \quad &::= \quad \ir{forall}({\ir{Index}}, \ir{ Stmt}) \\
\ir{\textcolor{\myred}{Forsome}} \quad &::= \quad \ir{\textcolor{\myred}{\text{forsome}({\ir{Index}}, \{\ir{Tensor*}\}, \ir{ Stmt})}} \\
\ir{\textcolor{\myred}{Forsame}} \quad &::= \quad \ir{\textcolor{\myred}{\text{forsame}({\ir{Index}}, \{\ir{Tensor*}\}, \ir{ Stmt})}} \\
\ir{Where} \quad &::= \quad \ir{Stmt where Stmt} \\
\ir{Access} \quad &::= \quad \ir{Tensor}(\ir{Indices}) \\
\ir{Indices} \quad &::= \quad \ir{Index*} \\
\ir{Expr} \quad &::= \quad \ir{Literal} \mid \ir{Access} \mid (\ir{Expr}) \mid \ir{Expr} \times \ir{Expr} \mid \dots
\end{align*}
\end{subfigure}
}
% \end{multicols}
\caption{Subset of the grammar of the extended TACO IR. The extensions are highlighted in \textcolor{\myred}{red}. In the diagram, the first letter of the non-terminals are capitalized.} %\ap{it's not great that the same words used as both non-terminals and terminals (forsome, forsame, where) and the differents in fonts is very hard to spot from the distance. The easiest way to fix it may be to capitalize non-terminals. Sometimes people use various sorts of special fonts, but they should be much more dissimilar then what is there now.}
\label{spconf:fig:grammar}
\end{figure*}

\begin{figure*}
\centering
\begin{subfigure}{0.15\textwidth}
\centering
\includegraphics[width=0.99\linewidth]{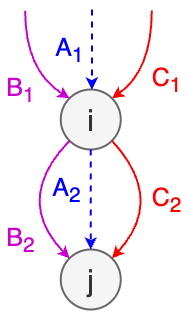}
\caption{}
\label{spconf:fig:hadamard-default-iter-graph}
\end{subfigure}
\hfill
\begin{subfigure}{0.20\textwidth}
\centering
\includegraphics[width=0.99\linewidth]{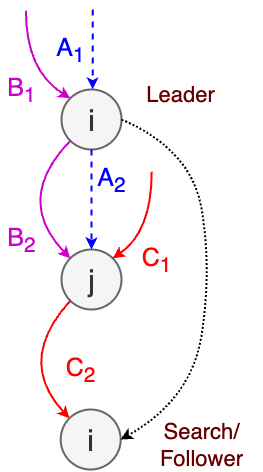}
\caption{}
\label{spconf:fig:hadamard-fused-iter-graph}
\end{subfigure}
\hfill
\begin{subfigure}{0.25\textwidth}
\begin{lstlisting}[basicstyle=\small, tabsize=2, showtabs=false, showstringspaces=false,morekeywords={forall,where},escapechar=\%]
forall i: 
  forall j: 
    A(i,j) = B(i,j) * C(i,j)
\end{lstlisting}
\caption{}
\label{spconf:fig:hadamard-default-ir}
\end{subfigure}
\hfill
\begin{subfigure}{0.27\textwidth}
\begin{lstlisting}[basicstyle=\small, tabsize=2, showtabs=false, showstringspaces=false,morekeywords={forall,where,forsome,forsame},escapechar=\%]
%\color{\myred}forsome% i %$\in\{A, B\}$%:
  forall j:
    %\color{\myred}forsame% i %$\in\{C\}$%:
      A(i,j) = B(i,j) * C(j,i)
\end{lstlisting}
\caption{}
\label{spconf:fig:hadamard-fused-ir}
\end{subfigure}
\caption{Iteration graphs and the corresponding IR of the Hadamard product. (a) Iteration graph of the Hadamard product when the indexing orders have the same order. (b) Iteration graph of the Hadamard product when the indexing orders have different orders. (c) IR of the Hadamard product when the indexing orders have the same order. (d) IR of the Hadamard product when the indexing orders have different orders.}
\label{spconf:fig:hadamard-ir}
\end{figure*}

\begin{figure*}[t]
\centering
\begin{minipage}{0.48\textwidth}
\begin{algorithm}[H]
\caption{\algcreateaccessmaplongform}
\label{spconf:alg:create-merge-lattice-points}
{\footnotesize
\begin{algorithmic}[1]
\Procedure{\algcreateaccessmap}{scs, expr}
    \State $\text{accessMap} \gets \text{Map(size=|scs|)}$ \Comment{ empty [] elements }
  
    \For{acc in \Call{TensorAccesses}{$expr$}}
        \State ptr $\gets 0$
        \For{idx in acc.indexVars}
            \If{\Call{isSparse}{acc}}
              \While{ptr $< |scs|$ \textbf{and} $scs[\text{ptr}] \neq \text{idx}$}
                  \State $\text{ptr} \gets \text{ptr} + 1$
              \EndWhile
            \Else \Comment{Dense tensor}
              \State $\text{ptr} \gets$ first position of idx in $scs$
            \EndIf
            \State $\text{accessMap}[\text{ptr}]$.append(acc)
        \EndFor
    \EndFor
    
    \State \textbf{return} accessMap
\EndProcedure
\end{algorithmic}
}
\end{algorithm}
\end{minipage}%
\hfill
\begin{minipage}{0.48\textwidth}
\begin{algorithm}[H]
\caption{\algconvertscstoirlongform}
\label{spconf:alg:convert-scs-to-ir}
{\footnotesize
\begin{algorithmic}[1]
\Procedure{\algconvertscstoir}{scs, accessMap, stmt}
    \State $\text{duplicateIndices} \gets$ \Call{DuplicateIndices}{scs}
    \State $\text{idxCntMap} \gets$ \Call{IdxCounts}{scs}
    
    \For{$i = |\text{scs}| - 1$ \textbf{down to} $0$} \Comment{Reverse order}
        \State $\text{idx} \gets \text{scs}[i]$
        \State $\text{accesses} \gets \text{accessMap}[i]$
        \If{$\text{idx} \notin \text{duplicateIndices}$}
            \State $\text{stmt} \gets$ \Call{Forall}{idx, stmt}
        \ElsIf{$\text{idxCntMap}[\text{idx}] > 1$}
            \State $\text{stmt} \gets$ \Call{Forsame}{idx, stmt, accesses}
            \State $\text{idxCntMap}[\text{idx}] \gets \text{idxCntMap}[\text{idx}] - 1$
        \Else
            \State $\text{stmt} \gets$ \Call{Forsome}{idx, stmt, accesses}
        \EndIf
    \EndFor
    
    \State \textbf{return} stmt
\EndProcedure
\end{algorithmic}
}
\end{algorithm}
\end{minipage}
\end{figure*}

\subsection{Internal Representation (IR)}\label{spconf:internal_representation}

We extend TACO's concrete index notation (CIN)~\cite{kjolstad:2018:workspaces}. Figure~\ref{spconf:fig:grammar} presents the grammar for this extended IR, with our additions highlighted in red. 

Two key components of the IR are the \texttt{Forall} and \texttt{Where} constructs. The \texttt{Forall} construct defines a loop over an index variable that iterates over all the tensors which contain that index, while the \texttt{Where} construct splits a computation into a producer and a consumer. The producer computes intermediate results into a dense temporary tensor, which the consumer then uses to generate the final output. In the grammar, \texttt{Index} and \texttt{Tensor} are identifiers, \texttt{Access} is a tensor with its indices, and \texttt{Expr} can be a literal, a tensor access, or a combination of expressions. 

To support the relaxed iteration graphs, we extend the IR with two new constructs. 

\begin{itemize}
    \item $\ir{Forsome}::= \texttt{forsome(Index, \{Tensor*\}, Stmt)}$: This construct defines a loop over the specified \texttt{index}, iterating only over the tensors listed in the set \{\texttt{Tensor*}\}. It is used when the index appears multiple times in the iteration graph, and we want to iterate over a subset of tensors that contain this index.
    \item $\ir{Forsame}::= \texttt{forsame(Index, \{Tensor*\}, Stmt)}$: This construct defines a search operation for the specified \texttt{index} within each of the indexing array of the tensors listed in \{\texttt{Tensor*}\} individually. It is used when the index appears multiple times in the iteration graph, and we want to find the same instance of this index as in a previous loop.
\end{itemize}

These two constructs allow us to represent the leader-follower iteration strategy in the IR, enabling efficient code generation for computations involving sparse tensors with differing index orders.

\begin{wrapfigure}{r}{0.4\textwidth}
  \centering
  \includegraphics[width=0.99\linewidth]{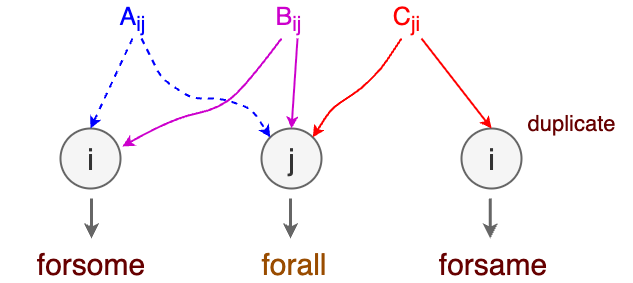}
  \caption{Example of IR generation from SCS.}
  \label{spconf:fig:ir-gen-example}
\end{wrapfigure}

The SCS is transformed into this IR. For the Hadamard product, Figures~\ref{spconf:fig:hadamard-default-iter-graph} and~\ref{spconf:fig:hadamard-default-ir} show the iteration graph and corresponding IR when the indexing orders are the same. When the orders differ, Figures~\ref{spconf:fig:hadamard-fused-iter-graph} and~\ref{spconf:fig:hadamard-fused-ir} illustrate the newly formed iteration graph and IR, respectively. 

The figures illustrate the use of the \texttt{forsome} and \texttt{forsame} iterators to define loops in the IR. The naming of these constructs reflects their function. The \texttt{forsome} iterator, a variation of TACO's \texttt{forall}, iterates over a subset of tensors containing a given index—in this case, iterating over index $i$ for tensors $\sparse B$ and $\sparse A$. Correspondingly, the \texttt{forsame} iterator is used to find the same instance of index $i$ from the \texttt{forsome} loop by searching within another tensor, here $\sparse C$.

The process of converting the SCS to our IR is illustrated in Figure~\ref{spconf:fig:ir-gen-example}. The algorithm begins by identifying duplicate indices within the SCS, which is the key to differentiating between \texttt{forall}, \texttt{forsome}, and \texttt{forsame} loop constructs. It then establishes a mapping from each index in the SCS to the set of tensor accesses that use that index. The construction of the SCS guarantees two properties: first, that every index from every tensor in the expression is present in the SCS, and second, that the SCS contains subsequences that respect the ordering constraints imposed by each sparse tensor. To assign tensor accesses to specific positions in the SCS, the algorithm proceeds from the start of each tensor's index list, advancing its position within the SCS for each index of a sparse tensor access.

We explain these procedures in Algorithms~\ref{spconf:alg:create-merge-lattice-points} and~\ref{spconf:alg:convert-scs-to-ir}. Algorithm~\ref{spconf:alg:create-merge-lattice-points} builds a map from each index in the SCS to its corresponding tensor accesses. This map is then consumed by Algorithm~\ref{spconf:alg:convert-scs-to-ir}, which traverses the SCS in reverse to generate the final IR, creating the appropriate loop constructs by checking for duplicated indices.

Our co-iteration\footnote{Co-iteration refers to iterating over indexing meta arrays of multiple tensors in tandem.} strategy is analogous to the merge lattice construction\footnote{Merge lattice construction is a technique used in TACO~\cite{taco} to efficiently coiterate over multiple sparse tensors by constructing a lattice of iteration points that represent the merged iteration space.} used in TACO, but with a key difference: we relax the requirement to co-iterate over all tensors in the contraction. This is achieved through our new iterators. The \texttt{forsome} iterator co-iterates only over the subset of tensors incident upon it, while the \texttt{forsame} iterator performs a targeted search operation on a single tensor.

\begin{figure*}
\centering
\begin{subfigure}{0.17\textwidth}
\includegraphics[width=0.99\linewidth]{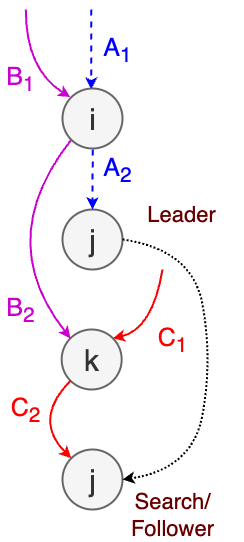}
\caption{}
\label{spconf:fig:spmm-fused-iteration-graph}
\end{subfigure}
\hfill
\begin{subfigure}{0.23\textwidth}
\includegraphics[width=0.99\linewidth]{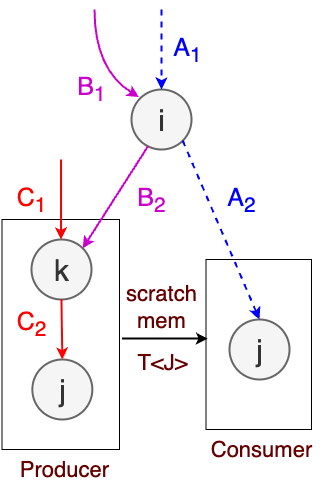}
\caption{}
\label{spconf:fig:spmm-branched-iteration-graph}
\end{subfigure}
\hfill
\begin{subfigure}{0.27\textwidth}
\begin{lstlisting}[basicstyle=\small, tabsize=2, showtabs=false, showstringspaces=false,morekeywords={forall,where,forsome,forsame},escapechar=\%]
forall i:
 %\color{\myred}forsome% j %$\in\{A\}$%:
  forall k:
   %\color{\myred}forsame% j %$\in\{C\}$%:
    A(i,j) += B(i,k) * C(k,j)
\end{lstlisting}
\caption{}
\label{spconf:fig:spmm-fused-ir}
\end{subfigure}
\hfill
\begin{subfigure}{0.26\textwidth}
\begin{lstlisting}[basicstyle=\small, tabsize=2, showtabs=false, showstringspaces=false,morekeywords={forall,where,forsome,forsame},escapechar=\%]
forall i:
 where:
  t(j) = 0 // workspace
  forall k: // producer
   forall j:
    t(j) += B(i,k) * C(k,j)
  forall j: // consumer
   A(i,j) = t(j)
\end{lstlisting}
\caption{}
\label{spconf:fig:spmm-branched-ir}
\end{subfigure}
\caption{(a) Relaxed iteration graph of SpMM. (b) Nested multi-inner/child iteration graph of SpMM. (c) IR of SpMM from the relaxed iteration graph (Figure~\ref{spconf:fig:spmm-fused-iteration-graph}). (d) IR from the Nested multi-inner/child iteration graph in Figure~\ref{spconf:fig:spmm-branched-iteration-graph}.}
\label{spconf:fig:spmm-ir}
\end{figure*}

\subsection{Transformations on the Relaxed Iteration Graph and the IR}\label{spconf:transformations_to_ir}

A significant advantage of having our relaxed iteration graph and the accompanying new IR constructs is the application of automatic transformations on this new IR to other already existing IR constructs in the TACO. This allows us to leverage existing optimizations and code generation techniques in TACO while supporting more flexible and efficient iteration patterns, all without requiring the user to manage the underlying complexity.

Now let's consider our second example of sparse matrix-sparse matrix multiplication (SpMM),
\[{\sparse A(i,j)} = \sum\nolimits_{k} {\sparse B(i,k)} * {\sparse C(k,j)}\]

Figure~\ref{spconf:fig:spmm-fused-iteration-graph} displays the relaxed iteration graph for SpMM, and Figure~\ref{spconf:fig:spmm-fused-ir} presents its corresponding IR. A key observation is that the \texttt{forsome} iterator is incident only upon the output tensor. Because the output tensor is sparse, we lack advance knowledge of which $j$ indices will be non-zero. A naive but correct implementation would therefore require the \texttt{forsome} iterator to traverse all possible values of $j$ and subsequently filter out those indices that did not actually contribute to the result after the contraction. This approach is not only computationally wasteful and inefficient but also adds significant implementation complexity. 

In these scenarios, we can transform the relaxed iteration graph into a more efficient branched loop nest structure~\cite{kjolstad:2018:workspaces, sparselnr, react}. This transformation is triggered by identifying loop indices that appear exclusively in the sparse output tensor and not in any of the input tensors. The result of this transformation is depicted in Figure~\ref{spconf:fig:spmm-branched-iteration-graph}, which shows the new iteration graph, and Figure~\ref{spconf:fig:spmm-branched-ir}, which shows the corresponding IR. This new structure splits the computation into two distinct phases using the \texttt{where} construct: a producer and a consumer. The producer computes the results into a dense temporary workspace, and the consumer then copies only the non-zero values from this workspace into the final sparse output tensor. This approach yields two significant benefits: it avoids the wasteful computation of iterating over all possible values of the $j$ index, and it allows us to leverage the powerful, pre-existing optimizations and code generation techniques available in TACO for these types of branched loop structures.

% An algorithm sketch of this transformation is shown in Algorithm~\ref{spconf:alg:transform-to-branched-ir}.

\begin{figure*}[t]
\centering
\begin{minipage}{\textwidth}
\begin{subfigure}{\textwidth}
\begin{equation*}
\frac{
    \begin{gathered}
    \ir{writes}(S_{\abody}) = \{A\} \quad \land \quad A \notin \ir{reads}(S_{\abody}) \\ %\text{\ap{what are ``writes'' and ``reads''?}}
    \ir{isSparse}(A(\vec{p})) = \mathrm{true} \quad \land \quad (A(\vec{p}) \mathrel{+}= E) \in S_{\abody} \\ %\text{\ap{what's this?}}
    i \in \ir{indices}(\vec{p}) \quad \land \quad \vec{p} = (\ldots, i) %\text{\ap{is this just $\vec{p} = (\ldots, i)$?}}
    \end{gathered}
}
{
    \tau[\mathit{op}(\ldots, \ir{forsome}(i, \{A\}, S_{\abody}))] \Rightarrow \mathit{op}(\ldots, \ir{forall}(i, A(\vec{p}) = t(i))\ \ \ir{where}\ \ \ir{rewrite}(i, S_{\abody}))
}
\end{equation*} 
\caption{Transformation rule. \ir{reads} returns the set of tensors in RHS of all \ir{Assignments}, \ir{writes} returns the set of all tensors in LHS of all \ir{Assignments}. $\mathit{op}$ is a \ir{Stmt}.} %\ap{if $t$ is a temporary, then it should probably be $T$, and it'd be great to introduce it above the line but I'm not sure if it's possible. What is op? is it a stmt?} \ad{didn't change to T because in the rewrite rule we use T to denote tensors.}
\label{spconf:eq:transformation-rule}
\end{subfigure}
\end{minipage}
\begin{minipage}{\textwidth}
\begin{subfigure}{\textwidth}
{\footnotesize
\begin{align*}
\ir{rewrite}(i, A(\vec{p}) \mathrel{+}= Expr) &\Rightarrow t(i) \mathrel{+}= Expr \\
\ir{rewrite}(i, \ir{forsame}(i, \{T*\}, S_{\inner})) &\Rightarrow
\begin{cases}
    \ir{forall}(i, \ir{rewrite}(i, S_{\inner})) & \text{if } |\ir{forsame}(i, \cdot) \text{ in } S_{\abody}| = 1 \\
    \ir{forsome}(i, \{T*\}, \ir{rewrite}(i, S_{\inner})) & \text{if first } \ir{forsame}(i, \cdot) \text{ in } S_{\abody} \\
    \ir{forsame}(i, \{T*\}, \ir{rewrite}(i, S_{\inner})) & \text{if subsequent } \ir{forsame}(i, \cdot) \text{ in } S_{\abody}
\end{cases} \\
\ir{rewrite}(i, \ir{op}(S_1, S_2, \ldots)) &\Rightarrow \ir{op}(\ir{rewrite}(i, S_1), \ir{rewrite}(i, S_2), \ldots)
\end{align*}
}
\caption{Helper function \texttt{rewrite($i, S$)} to rewrite the producer part. This function is applied recursively to the statement $S$.}
\label{spconf:eq:helper-function}
\end{subfigure}
\end{minipage}
\caption{Transformation $\tau$ from a \texttt{forsome} iterator to a \texttt{where} clause with a producer and consumer part. $A$ is the output tensor, $i$ is an iterator index variable, $S, S_{body}, S_{inner}$ are \ir{Stmt}s, $\vec{p}$ is a vector of index variables, and $t$ is a fresh temporary dense workspace tensor.} %\ap{It looks like rewrite implicitly depends on $i$, and this is bad. But it's easy to fix: it should have a parameter for $i$, i.e. instead of rewrite(S) it should be rewrite(i, S)} \ap{the caption sounds very specific. Somewhere there should be an explanation that this transformation is applied recursively to the whole ``program''.}
\label{spconf:eq:transformation-rules}
\end{figure*}

We describe this transformation formally using transformation rules in Figure~\ref{spconf:eq:transformation-rules}. 
% We use the notation below to describe the rule:
% \begin{itemize}
%   \item $A$: the output tensor.
%   \item $i$: the sparse iterator index variable.
%   \item $S, S_{body}, S_{inner}$: statements or statement blocks in the IR.
%   \item $E$: an expression in the IR. \ap{you shouldn't need to re-introduce parts of syntax that you already introduced on the FIgure with syntax (Fig 6, I think). You only need to be consistent with the Fig, so, for instance, instead of E it should be expr, and instead of S it should be stmt}
%   \item $\vec{p}$: a vector of index variables.
%   \item $t$: a fresh temporary dense workspace tensor.
% \end{itemize}
% \ap{ideally, all these notations should be in the caption or in the figure itself: it's very confusing to see things like $t$ there and only later discover this list of notations}

% \ap{is this paragraph describing Figure 10? it should say so (probably, just move the last sentence of the section here). You could label certain parts of the figure and refer to the labels (e.g. the three rules for rewrite could be a numbered list)} \ad{fixed as exactly requested.}

The transformation rule $\tau$ is defined as in Figure~\ref{spconf:eq:transformation-rule}. Our implementation of this transformation begins by identifying an outermost \texttt{forsome} iterator that exclusively iterates over the output tensor. Once identified, we rewrite the IR by splitting the computation into a producer and a consumer, encapsulated within a \texttt{where} construct. The consumer part, $\ir{forall}(i, A(\vec{p}) = t(i))$, is created as a new loop that iterates over the original \texttt{forsome} index, copying the non-zero values from the dense workspace to the sparse output tensor. The producer, $\ir{rewrite}(i, S_{\abody})$, is constructed by duplicating the loop nest originally inside the identified \texttt{forsome} iterator. Finally, the original \texttt{forsome} iterator is removed from the IR. 

The \texttt{rewrite(S)} helper function, detailed in Figure~\ref{spconf:eq:helper-function}, applies a set of transformations to the producer block. These transformations consist of the following rules:
\begin{enumerate}
    \item Accesses to the output tensor are redirected to a dense temporary workspace.
    \item The first \texttt{forsame} iterator within the producer section that corresponds to the \texttt{forsome} iterator index is converted. If it is the only such iterator, it becomes a \texttt{forall} iterator; otherwise, it is rewritten as a \texttt{forsome} iterator.
    \item All other statements are recursively processed to apply the same transformation rules.
\end{enumerate}
This process yields a more efficient computation by eliminating unnecessary iterations over non-contributing indices, while still benefiting from the existing optimization and code generation capabilities within TACO.

\section{IMPLEMENTATION}\label{spconf:implementation}

We use TACO as the base library for our implementation. We build our library in C++ and use TACO's formats, tensor reading, and other functionalities. 

We implement our relaxed iteration graph construction (\S~\ref{spconf:relaxed_iteration_graph}) by modifying TACO's existing iteration graph construction code. We extend TACO's concrete index notation (CIN)~\cite{kjolstad:2018:workspaces} to introduce the new IR nodes \texttt{forsome} and \texttt{forsame} (\S~\ref{spconf:internal_representation}). We implement the code generation for these new IR nodes by modifying TACO's existing code generation module. We also implement the transformation rules to convert the new IR into existing IR constructs (\S~\ref{spconf:transformations_to_ir}) by adding a new transformation pass in TACO's existing transformation framework. We limit this transformation pass to be applied only when the new IR can be converted to a branched IR with only one 1-D intermediate temporary tensor. We impose this restriction because 2-D dense intermediate temporaries are usually large leading to high memory consumption, and slower performance. We assume all the storage formats are ordered and unique, and we only add support for Compressed Sparse Fiber (CSF) like formats and dense formats.

\section{EVALUATION}\label{spconf:evaluation}

This section evaluates \system against TACO and other state-of-the-art tensor transpose libraries through two case studies. In the first study (\S~\ref{spconf:case_study_1}), we assess the performance of \system on tensor contractions that require searching over the indexing arrays of input tensors. In the second study (\S~\ref{spconf:case_study_2}), we evaluate its ability to generate code for contractions that need to be reordered into a branched loop nest.

\subsection{Experimental Setup}

All experiments were conducted on a single-socket 64-Core AMD Ryzen Threadripper 3990X processor clocked at 2.2 GHz. The machine is equipped with a 32 KB L1 data cache, a 512 KB shared L2 cache, and a 16 MB shared L3 cache. We compiled all code using GCC 11.4.0 with the \texttt{-O3} and \texttt{-ffast-math} optimization flags. For each benchmark, we report the median of 31 serial executions to ensure stable and reliable performance measurements.

\begin{table*}[t]\centering
\caption{Tensors and matrices used in the evaluation from various matrix and tensor collections}
\label{spconf:tab:datasets}
% \vspace{-0.5em}
% \hspace{-4.0em}
\begin{adjustbox}{minipage=\linewidth,scale=0.99}
\centering
\footnotesize
% \begin{center}
\begin{tabular}{lrrr} % <-- Alignments: 1st column left, 2nd middle and 3rd right, with vertical lines in between
\toprule
\textbf{Tensor} & \textbf{Dimensions} & \textbf{Non-zeros} & \textbf{Sparsity}\\
\midrule
bcsstk17 & { $11K \times 11K$} & {$429K$} & $4\mathrm{e}{-3}$\\
scircuit & { $171K \times 171K$} & {$959K$} & $3\mathrm{e}{-5}$\\
mac\_econ\_fwd500 & { $207K \times 207K$} & {$1.27M$} & $9\mathrm{e}{-5}$\\
majorbasis & { $160K \times 160K$} & {$1.75M$} & $7\mathrm{e}{-5}$\\
Lin & { $256K \times 256$} & {$1.77M$} & $3\mathrm{e}{-5}$\\
rma10 & { $47K \times 47K$} & {$2.37M$} & $1\mathrm{e}{-3}$\\
cop20k\_A & { $12K \times 12K$} & {$2.62M$} & $2\mathrm{e}{-4}$\\
webbase-1M & { $1.00M \times 1.00M$} & {$3.11M$} & $3\mathrm{e}{-6}$\\
cant & { $62K \times 62K$} & {$4.01M$} & $1\mathrm{e}{-3}$\\
pdb1HYS & { $36K \times 36K$} & {$4.34M$} & $3\mathrm{e}{-3}$\\
ecology1 & { $1M \times 1M$} & {$5.00M$} & $5\mathrm{e}{-6}$\\
largebasis & {$440K \times 440K$} & {$5.24M$} & $2.7\mathrm{e}{-5}$\\
consph & { $83K \times 83K$} & {$6.01M$} & $9\mathrm{e}{-4}$\\
shipsec1 & { $140K \times 140K$ } & {$7.81M$} & $2\mathrm{e}{-4}$\\
atmosmodd & { $1.27M \times 1.27M$ } & {$8.81M$} & $5.4\mathrm{e}{-6}$\\
pwtk & { $217K \times 217K$} & {$11.52M$} & $2.4\mathrm{e}{-4}$\\
\midrule%
% delicious-3d & {\footnotesize $532924 \times 17262471 \times 532924$} & {\footnotesize $140126181$}\\
vast-2015-mc1-3d & { $165K \times 11K \times 2$} & { $26.02M$} & $8.36\mathrm{e}{-08}$\\
darpa1998 & { $22K \times 22K \times 23.7M$} & { $28.42M$} & $2.50\mathrm{e}{-06}$\\
nell-2 & { $12K \times 9K \times 288K$} & { $76.88M$} & $5.73\mathrm{e}{-05}$\\
freebase\_music & { $23M \times 23M \times 23M$} & { $99.55M$} & {$8.13\mathrm{e}{-15}$}\\
flickr-3d & { $320K \times 2.82M \times 1.60M$} & { $112.89M$} & $3.92\mathrm{e}{-11}$\\
freebase\_sampled & {$39M \times 39M \times 39M$} & {$139.92M$} & {$2.36\mathrm{e}{-15}$} \\
nell-1 & { $ 2.9M \times 2.1M \times 25.5M$} & { $143.60M$} & {$9.27\mathrm{e}{-13}$}\\
\bottomrule
\end{tabular}
% \end{center}
\end{adjustbox}
% \vspace{-1.0em}
\end{table*}

\subsection{Datasets}\label{spconf:datasets}

Our evaluation uses a diverse set of real-world tensors from the SuiteSparse Collection~\cite{suitesparse}, Network Repository~\cite{networkrepository}, Formidable Repository of Open Sparse Tensors and Tools (FROSTT)~\cite{frosttdataset}, and the 1998 DARPA Intrusion Detection Evaluation Dataset~\cite{darpa}. Table~\ref{spconf:tab:datasets} details these tensors, which cover a wide range of sizes and sparsities. We represent two-dimensional sparse matrices in the Compressed Sparse Row (CSR) format and three-dimensional sparse tensors in the Compressed Sparse Fiber (CSF) format. To ensure matching dimensions for tensor contractions, we duplicate input tensors when necessary.

\begin{figure*}[!t]
%\begin{subcaptiongroup}
\centering
\begin{subfigure}{0.49\textwidth}
    \centering
    \includegraphics[width=.95\linewidth]{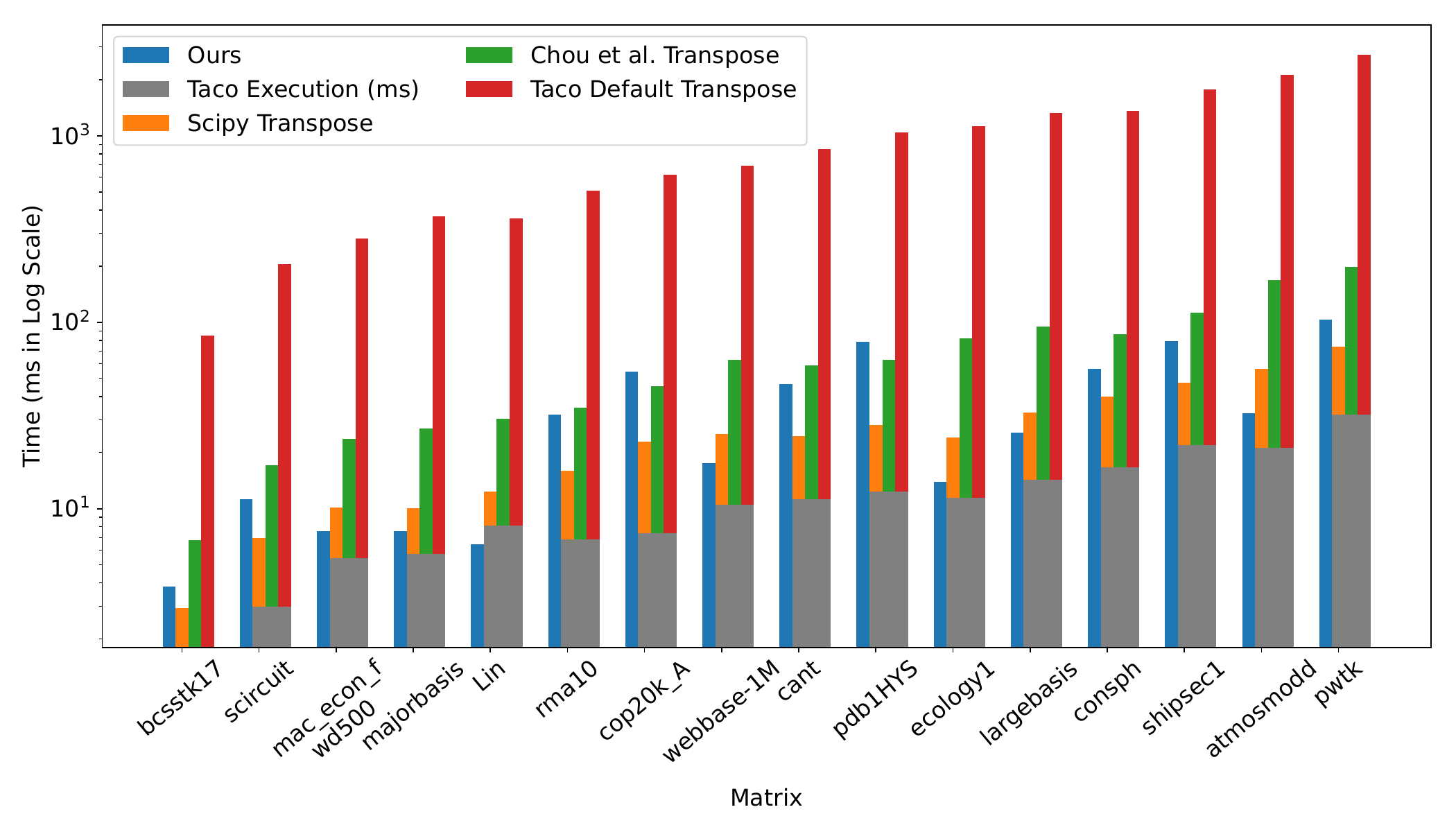}
    \caption{Execution time.}
    \label{spconf:fig:elmultiplication-noassem-1}
\end{subfigure}%
\begin{subfigure}{0.49\textwidth}
    \centering
    \includegraphics[width=.95\linewidth]{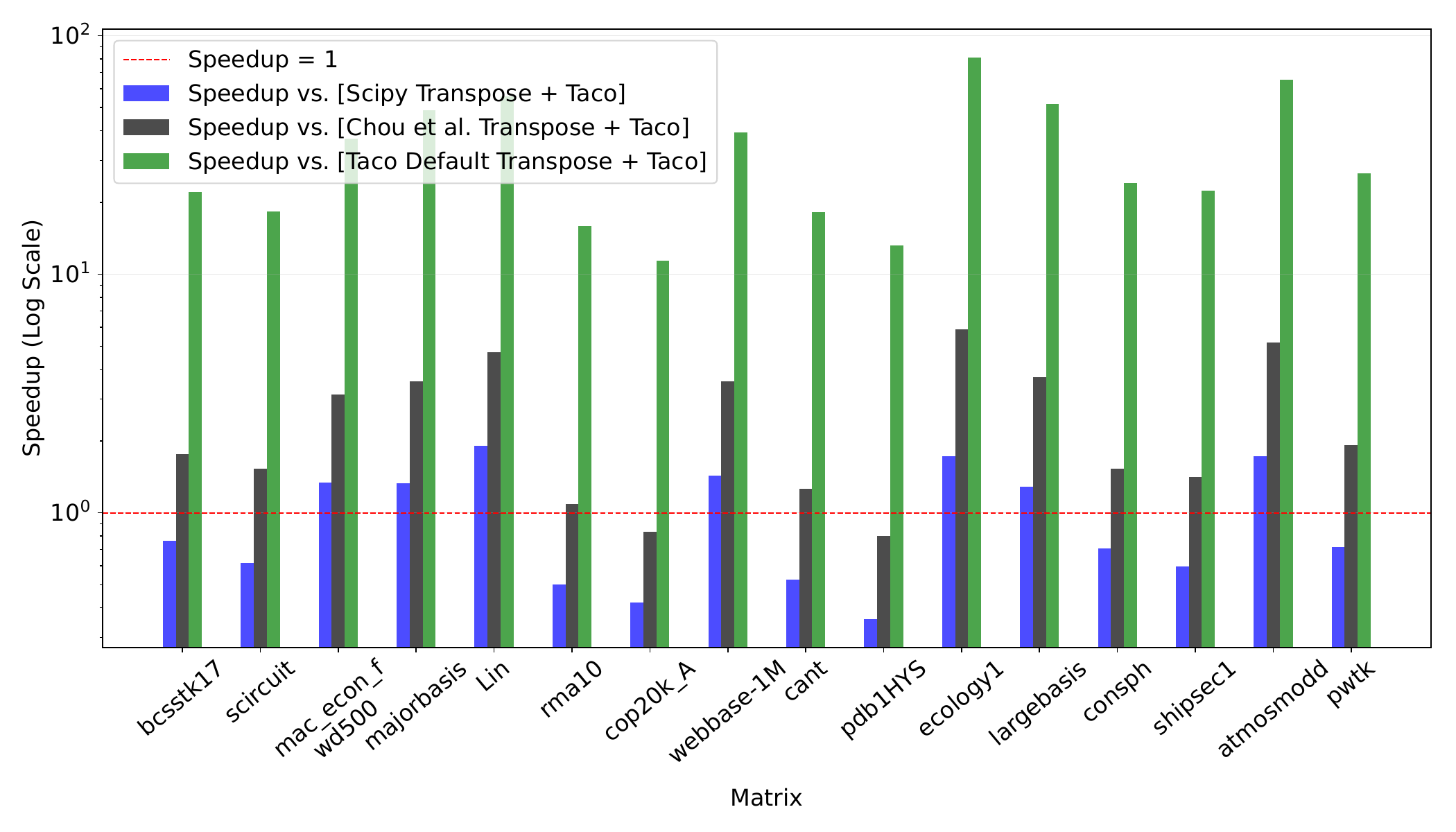}
    \caption{Speedup (Baseline Time / \system Time).}
    \label{spconf:fig:elmultiplication-noassem-2}
\end{subfigure}%
\caption{Performance evaluation of elementwise multiplication kernel $A(i,j) = B(i,j) * C(j,i)$ without output index array assembly during compute.}
\label{spconf:fig:elmultiplication-noassem}
\end{figure*}

\begin{figure*}[t]
\begin{subfigure}{0.49\textwidth}
    \centering
    \includegraphics[width=.95\linewidth]{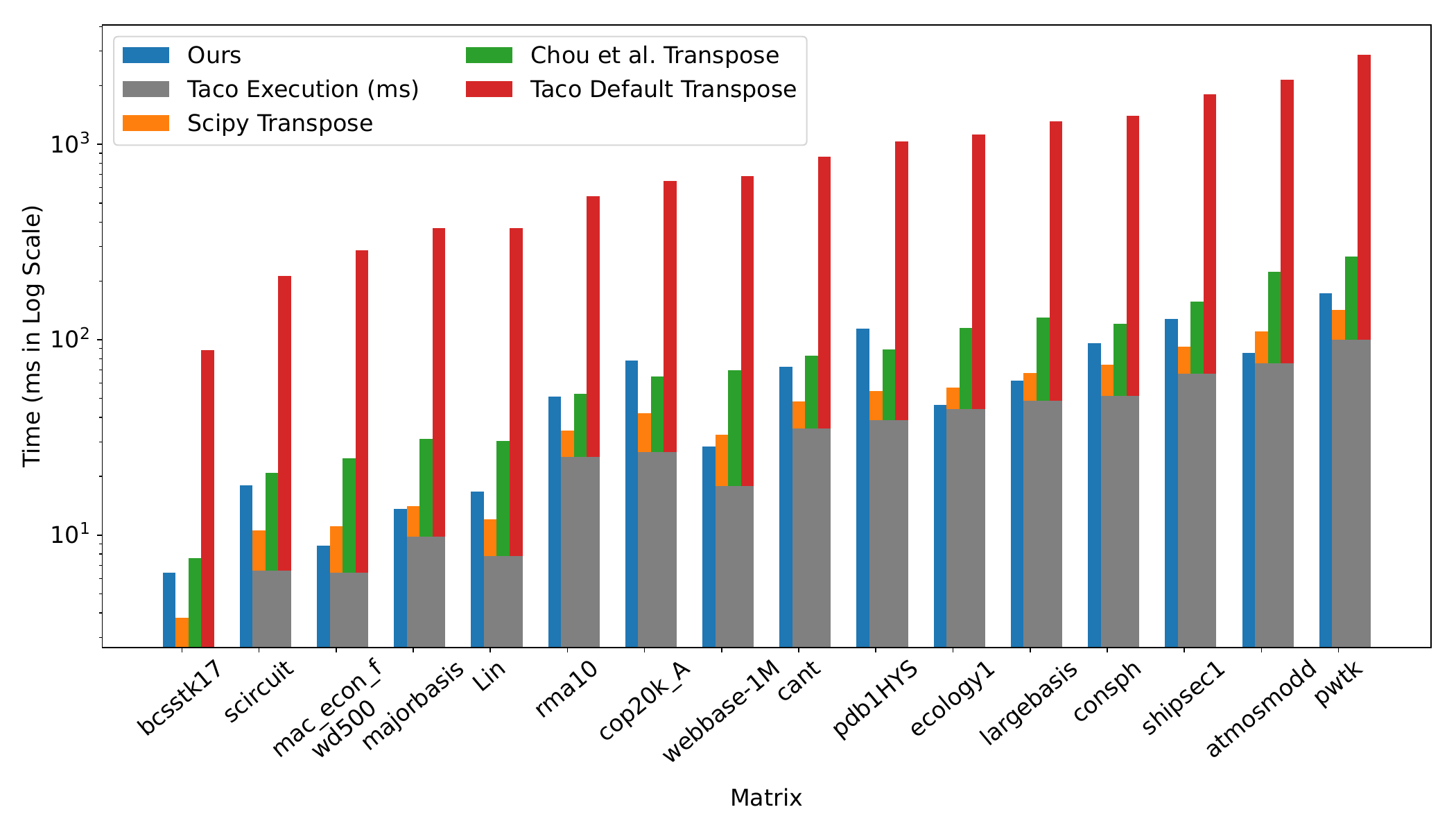}
    \caption{Execution time.}
    \label{spconf:fig:elmultiplication-wassem-1}
\end{subfigure}
\begin{subfigure}{0.49\textwidth}
    \centering
    \includegraphics[width=.95\linewidth]{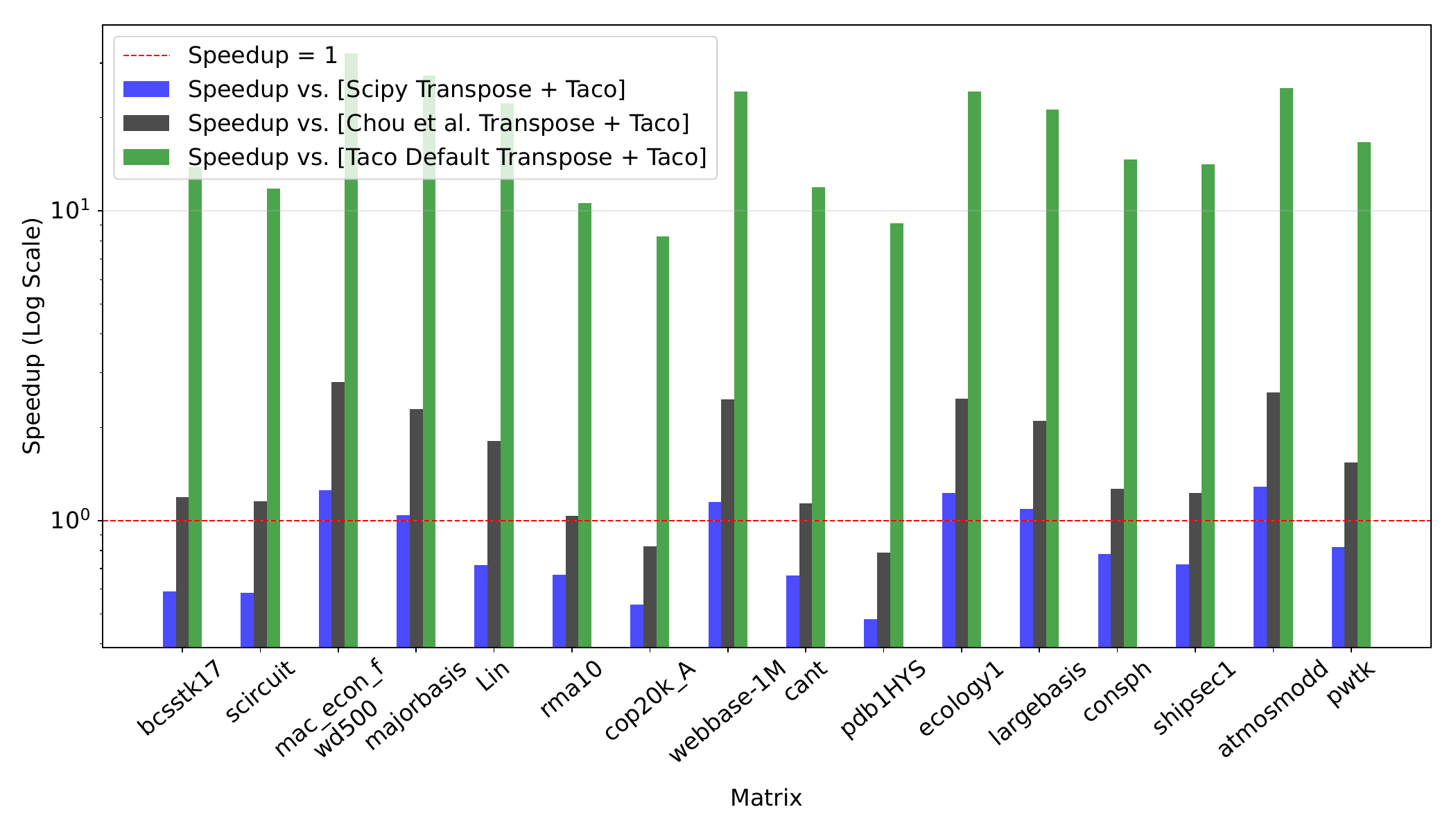}
    \caption{Speedup (Baseline Time / \system Time).}
    \label{spconf:fig:elmultiplication-wassem-2}
\end{subfigure}
\caption{Performance evaluation of elementwise multiplication kernel $A(i,j) = B(i,j) * C(j,i)$ with output index array assembly during compute.}
\label{spconf:fig:elmultiplication-wassem}
\end{figure*}

\begin{figure*}[t]
\centering
\begin{subfigure}{0.20\textwidth}
    \centering
    \includegraphics[width=\linewidth,trim=32mm 15mm 32mm 12mm,clip]{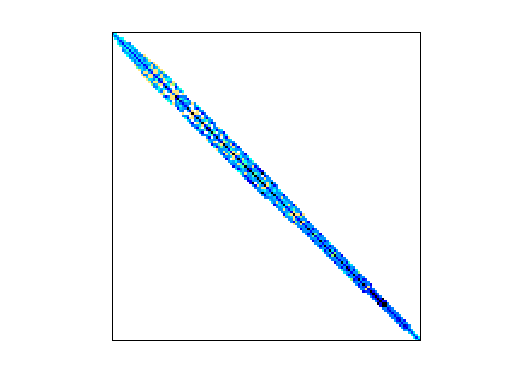}
    \caption{$shipsec1$}
    \label{spconf:fig:shipsec1}
\end{subfigure}%
\begin{subfigure}{0.20\textwidth}
    \centering
    \includegraphics[width=\linewidth,trim=32mm 15mm 32mm 12mm,clip]{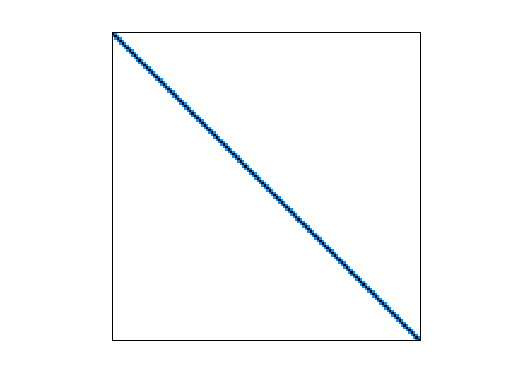}
    \caption{$ecology1$}
    \label{spconf:fig:ecology1}
\end{subfigure}%
\begin{subfigure}{0.20\textwidth}
    \centering
    \includegraphics[width=\linewidth,trim=32mm 15mm 32mm 12mm,clip]{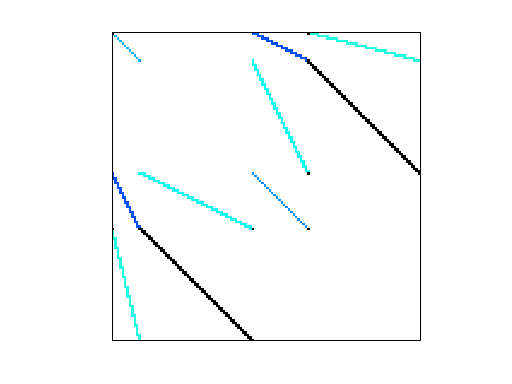}
    \caption{$largebasis$}
    \label{spconf:fig:largebasis}
\end{subfigure}%
\begin{subfigure}{0.20\textwidth}
    \centering
    \includegraphics[width=\linewidth,trim=32mm 15mm 32mm 12mm,clip]{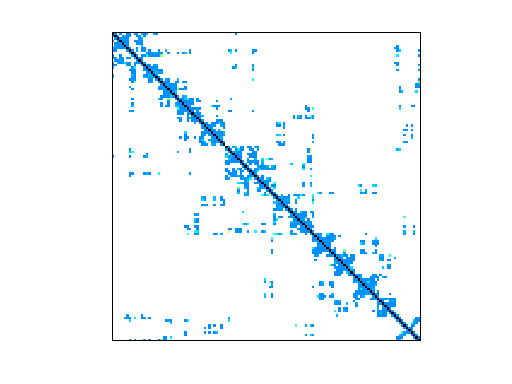}
    \caption{$pdb1HYS$}
    \label{spconf:fig:pdb1HYS}
\end{subfigure}%
\begin{subfigure}{0.20\textwidth}
    \centering
    \includegraphics[width=\linewidth,trim=32mm 15mm 32mm 12mm,clip]{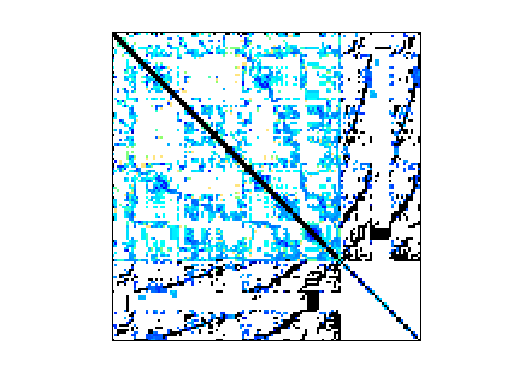}
    \caption{$scircuit$}
    \label{spconf:fig:scircuit}
\end{subfigure}%
\caption{Sparsity patterns of the matrices used in the elementwise multiplication benchmarks.}
\label{spconf:fig:matrices}
\end{figure*}

\begin{figure*}[t]
\centering
\begin{subfigure}{0.30\textwidth}
    \centering
    \includegraphics[width=\linewidth]{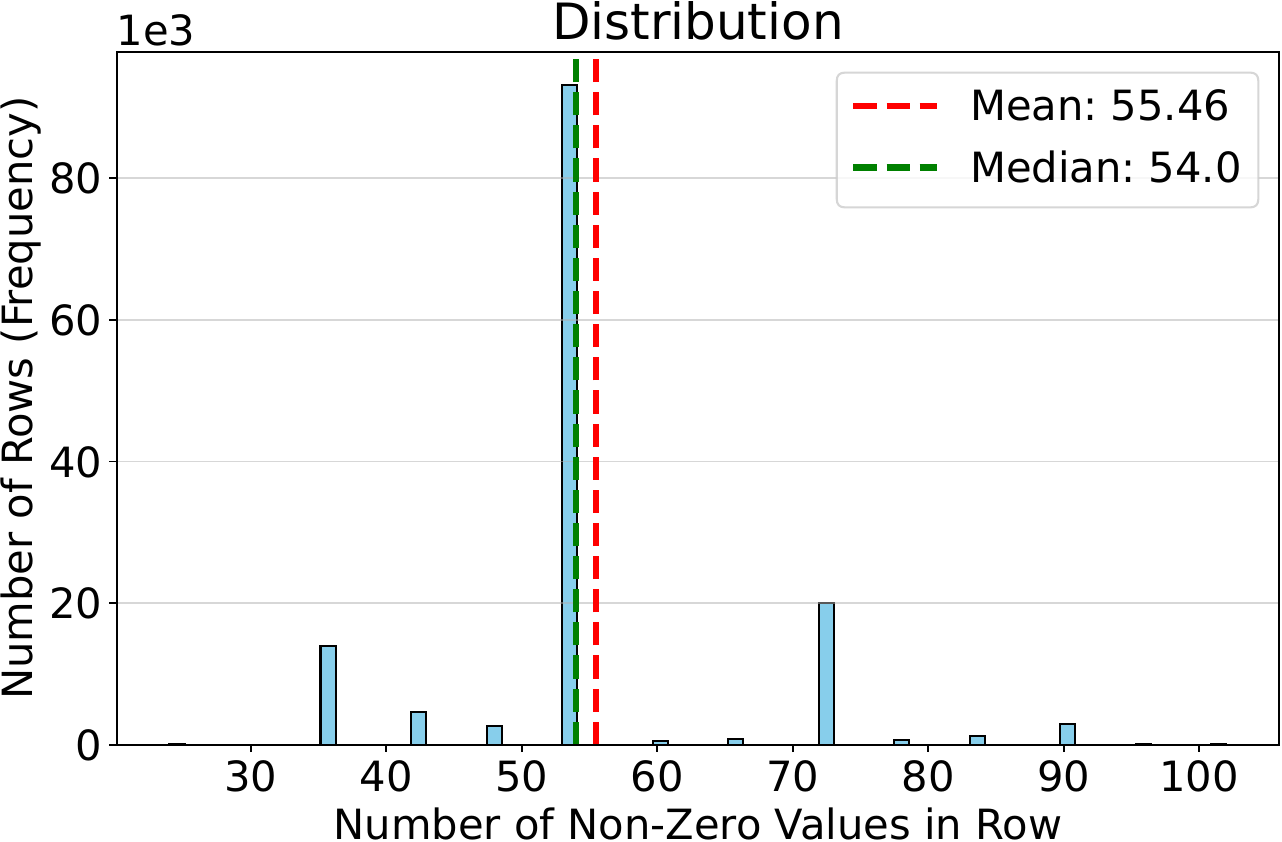}
    \caption{\textit{shipsec1}}
    \label{spconf:fig:shipsec1-histo}
\end{subfigure}%
\hspace{1em}
\begin{subfigure}{0.30\textwidth}
    \centering
    \includegraphics[width=\linewidth]{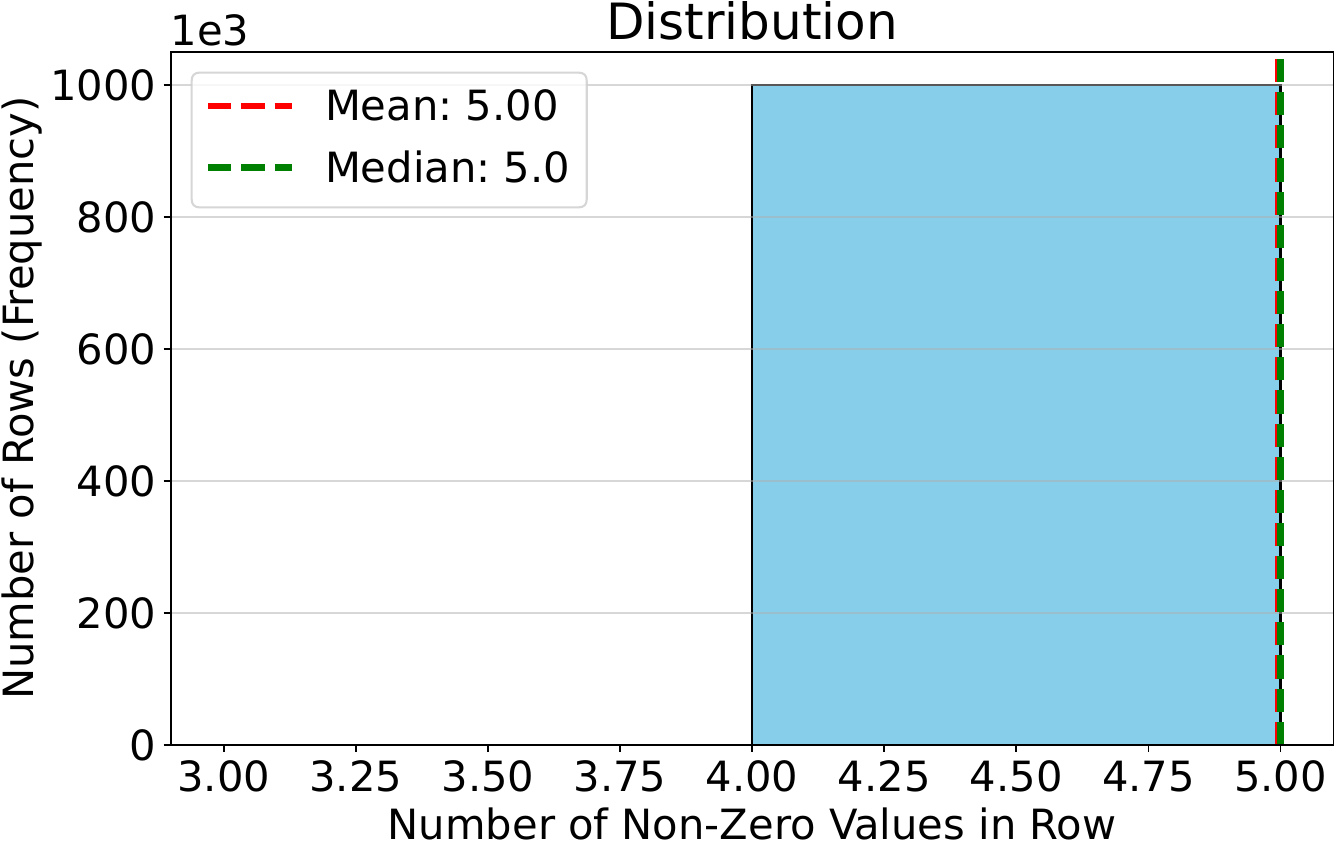}
    \caption{\textit{ecology1}}
    \label{spconf:fig:ecology1-histo}
\end{subfigure}%
\hspace{1em}
\begin{subfigure}{0.30\textwidth}
    \centering
    \includegraphics[width=\linewidth]{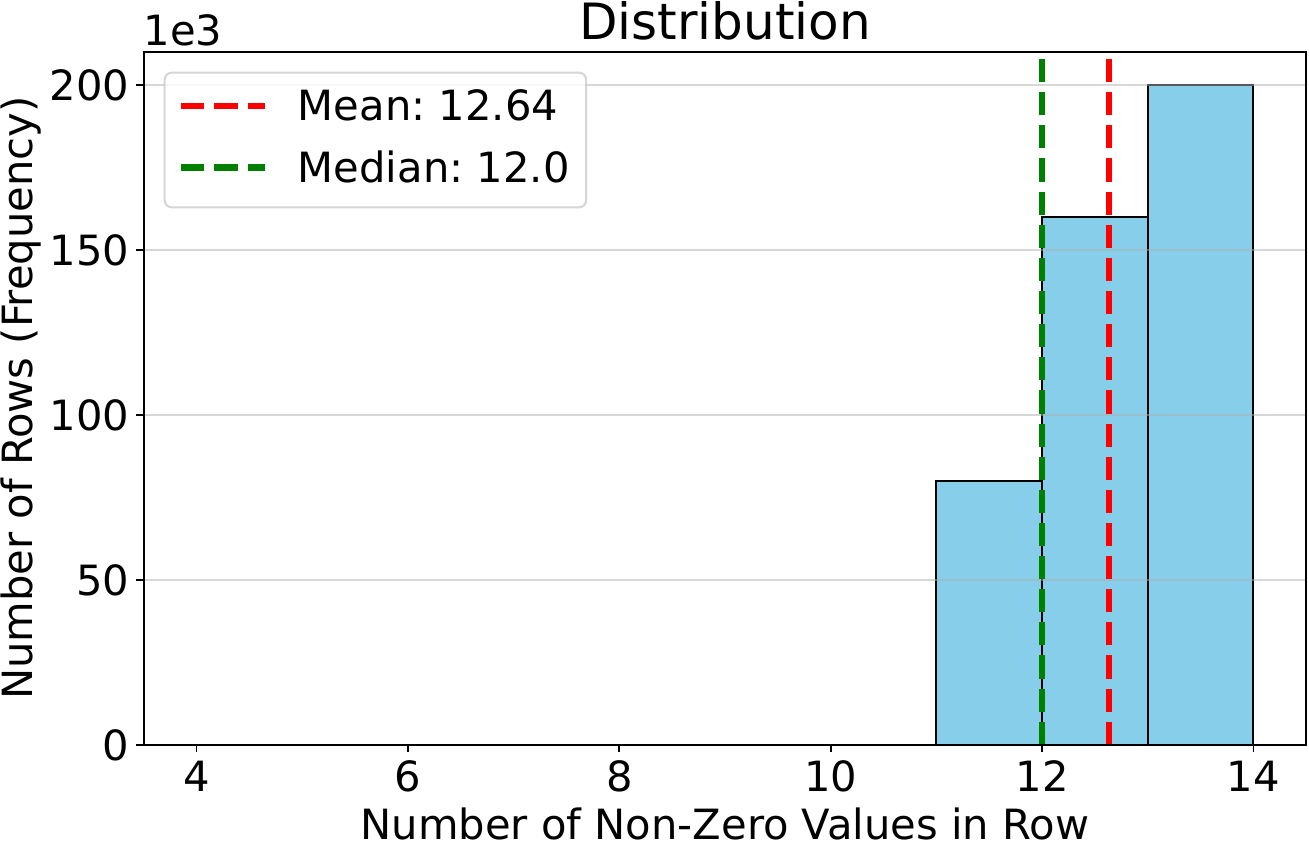}
    \caption{\textit{largebasis}}
    \label{spconf:fig:largebasis-histo}
\end{subfigure}%
\vskip\baselineskip
\begin{subfigure}{0.30\textwidth}
    \centering
    \includegraphics[width=\linewidth]{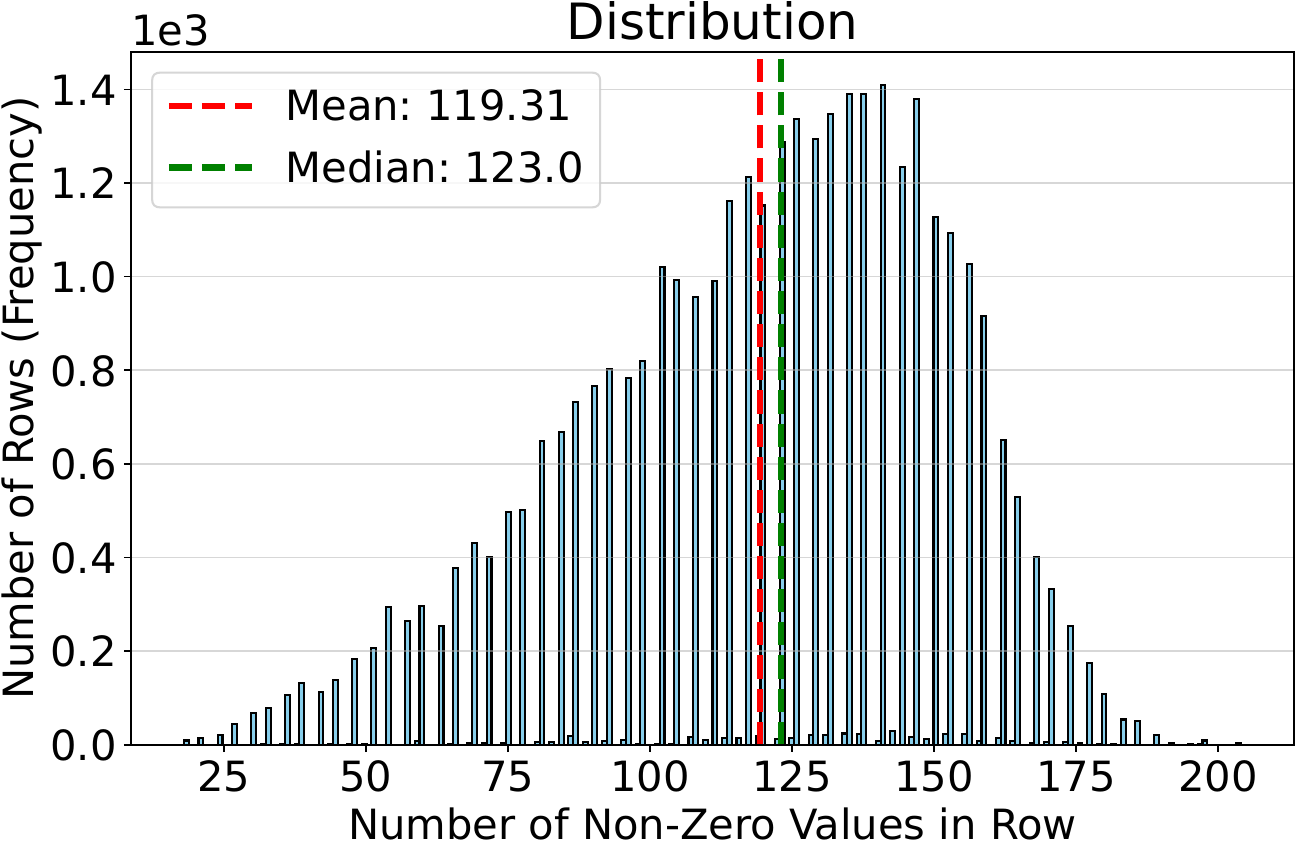}
    \caption{\textit{pdb1HYS}}
    \label{spconf:fig:pdb1HYS-histo}
\end{subfigure}%
\hspace{2em}
\begin{subfigure}{0.30\textwidth}
    \centering
    \includegraphics[width=\linewidth]{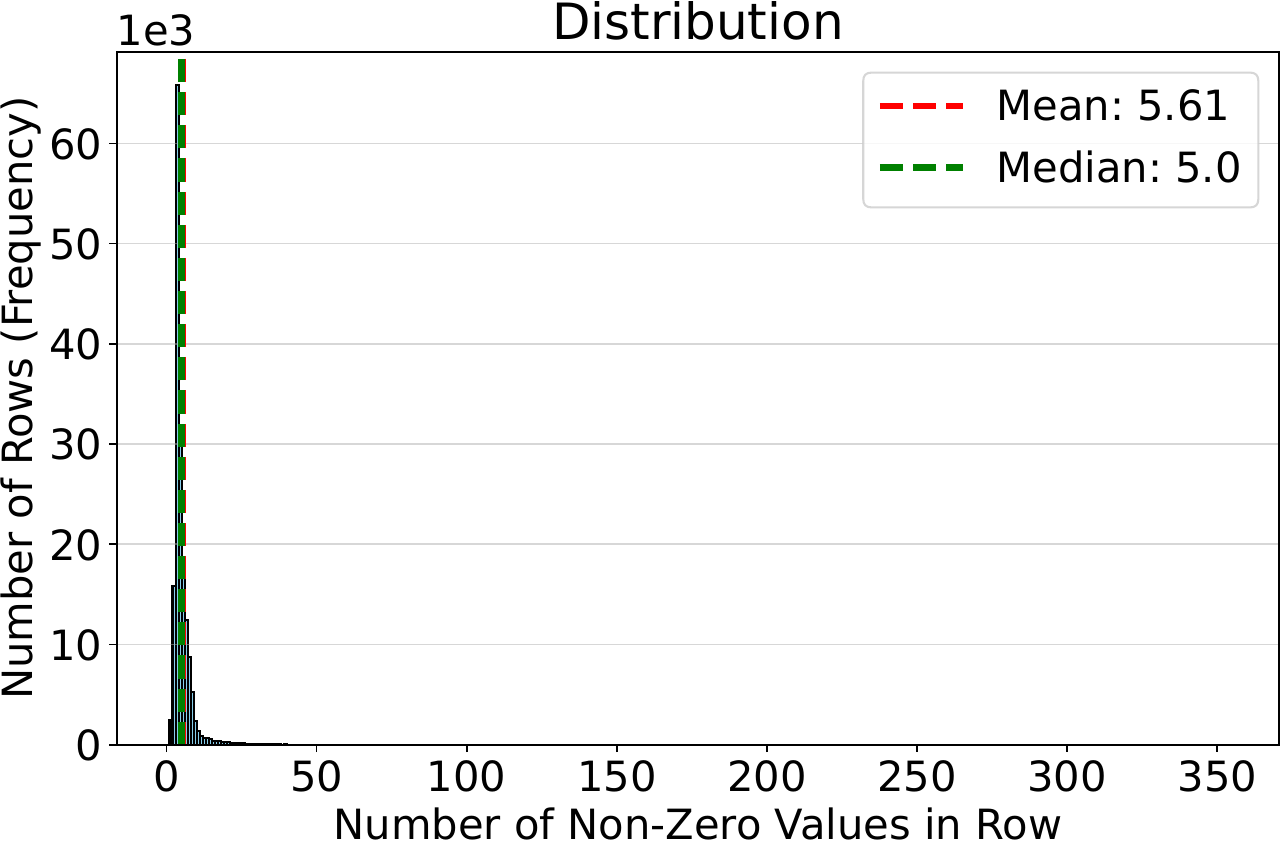}
    \caption{\textit{scircuit}}
    \label{spconf:fig:scircuit-histo}
\end{subfigure}%
\caption{Histograms of the number of non-zero elements in rows of the matrices.}
\label{spconf:fig:matrices-histo}
\end{figure*}

\subsection{Case Study 1: Tensor contractions containing search over indexing arrays}\label{spconf:case_study_1}

The first set of experiments evaluates \system's performance on tensor contractions that require searching over an input tensor's indexing arrays. This technique is particularly useful when one of the input tensors must be transposed to match the computation's requirements.

For this case study, we evaluate the performance of the following kernels:
\begin{description}
    \item[Element-wise-mul] An element-wise product of two matrices. \\${\sparse A(i,j)} = {\sparse B(i,j)} * {\sparse C(j,i)}$
    \item[Dot-product] A batched dot product between the rows of one matrix and the columns of another. \\$a(i) = {\sparse B(i,j)} * {\sparse C(j,i)}$
    \item[3D-tensor-contraction] A contraction of two 3D tensors. \\$a(i) = {\sparse B(i,j,k)} * {\sparse C(i,k,j)}$
\end{description}

We present the results for the element-wise multiplication kernel in Figures~\ref{spconf:fig:elmultiplication-noassem} and \ref{spconf:fig:elmultiplication-wassem}. We compare the performance of \system against baselines that adopt a two-step, transpose-then-compute strategy. First, one of the input tensors is explicitly transposed. Second, the kernel is executed using the pre-transposed tensor. All baseline computations are performed using TACO's default schedule. We evaluate the following transpose methods for the baselines:

\begin{description} 
    
\item[TACO Transpose] We use TACO's built-in functionality to perform the initial tensor transposition. This approach combines TACO's transpose with its default computation schedule. 

\item[SOTA Transpose] We use other state-of-the-art libraries for the transposition. For 2D sparse matrices, we use \texttt{SciPy.sparse}. For 3D sparse tensors in COO format, we measure the time taken by the C standard library implementation of \texttt{quick sort} function to reorder the indices, representing a lower bound on the achievable performance for 3D tensor transposition. 

\end{description}

Since the output of this element-wise multiplication is also a sparse matrix, we evaluate two scenarios. In the first, presented in Figure~\ref{spconf:fig:elmultiplication-noassem}, we generate the output matrix indexing arrays first and then report only the time taken to compute the output matrix value array. In the second, shown in Figure~\ref{spconf:fig:elmultiplication-wassem}, we generate the output matrix indexing arrays while computing the output values. The first case is useful for computations where the output structure is known \textit{a priori}, and only the output values need to be computed. In this computation, we use a duplicate copy of the same input matrix for both inputs to ensure matching dimensions.

The results are similar for both scenarios, although the second takes longer due to the overhead of assembling the output matrix indexing arrays concurrently with the output values. In both cases, \system outperforms the [TACO Transpose + TACO compute] baseline by a large margin. \system is also faster than TACO in most cases when considering the [TACO (\cite{chou2020-format-conversion} Transpose) + TACO compute] baseline. However, when a SOTA transpose library is used, \system is faster in only some cases.

The performance of \system relative to the baselines depends on the input matrices' sparsity patterns (Figure~\ref{spconf:fig:matrices}). For instance, with \textit{shipsec1} (Figure~\ref{spconf:fig:shipsec1}), \textit{ecology1} (Figure~\ref{spconf:fig:ecology1}), and \textit{largebasis} (Figure~\ref{spconf:fig:largebasis}), \system is faster than all baselines. However, for \textit{pdb1HYS} (Figure~\ref{spconf:fig:pdb1HYS}) and \textit{scircuit} (Figure~\ref{spconf:fig:scircuit}), it is slower than the fastest baseline. This performance difference arises because the non-zero elements in \textit{shipsec1} and \textit{ecology1} are clustered around the diagonal or have some structure, whereas in \textit{pdb1HYS} and \textit{scircuit}, they are scattered across the matrix.

\begin{comment}
min_row_nnz, max_row_nnz, avg_row_nnz, median_row_nnz, std_row_nnz
bcsstk17, 1.00, 150.00, 45.00, 39.06, 15.41
scircuit, 1.00, 353.00, 5.00, 5.61, 4.39
mac_econ_fwd500, 1.00, 44.00, 5.00, 6.17, 4.44
majorbasis, 6.00, 11.00, 11.00, 10.94, 0.42
Lin, 4.00, 7.00, 7.00, 6.90, 0.31
rma10, 4.00, 145.00, 46.00, 50.69, 27.78
cop20k_A, 0.00, 81.00, 23.00, 21.65, 13.79
webbase-1M, 1.00, 4700.00, 2.00, 3.11, 25.35
cant, 1.00, 78.00, 74.00, 64.17, 14.06
pdb1HYS, 18.00, 204.00, 123.00, 119.31, 31.86
ecology1, 3.00, 5.00, 5.00, 5.00, 0.06
largebasis, 4.00, 14.00, 12.00, 12.64, 1.15
consph, 1.00, 81.00, 81.00, 72.13, 19.08
shipsec1, 24.00, 102.00, 54.00, 55.46, 11.07
atmosmodd, 4.00, 7.00, 7.00, 6.94, 0.24
pwtk, 2.00, 180.00, 54.00, 53.39, 4.74
\end{comment}

\begin{table*}[t]
\caption{Row-statistics of the matrices used in our evaluation.}
\label{spconf:tab:matrix-stats}
\begin{minipage}{.5\linewidth}
\centering
\small
\begin{tabular}{lrrrrr}
\toprule
\textbf{Matrix} & \textbf{\scriptsize Min} & \textbf{\scriptsize Max} & \textbf{\scriptsize Med.} & \textbf{\scriptsize Mean} & \textbf{\scriptsize Std. Dev.}\\
\midrule
bcsstk17 & 1 & 150 & 45 & 39.06 & 15.41\\
scircuit & 1 & 353 & 5 & 5.61 & 4.39\\
mac\_econ\_f & 1 & 44 & 5 & 6.17 & 4.44\\
majorbasis & 6 & 11 & 11 & 10.94 & 0.42\\
Lin & 4 & 7 & 7 & 6.90 & 0.31\\
rma10 & 4 & 145 & 46 & 50.69 & 27.78\\
cop20k\_A & 0 & 81 & 23 & 21.65 & 13.79\\
webbase-1M & 1 & 4700 & 2 & 3.11 & 25.35\\
\bottomrule
\end{tabular}
\end{minipage}%
\begin{minipage}{.5\linewidth}
\centering
\small
\begin{tabular}{lrrrrr}
\toprule
\textbf{Matrix} & \textbf{\scriptsize Min} & \textbf{\scriptsize Max} & \textbf{\scriptsize Med.} & \textbf{\scriptsize Mean} & \textbf{\scriptsize Std. Dev.}\\
\midrule
cant & 1 & 78 & 74 & 64.17 & 14.06\\
pdb1HYS & 18 & 204 & 123 & 119.31 & 31.86\\
ecology1 & 3 & 5 & 5 & 5 & 0.06\\
largebasis & 4 & 14 & 12 & 12.64 & 1.15\\
consph & 1 & 81 & 81 & 72.13 & 19.08\\
shipsec1 & 24 & 102 & 54 & 55.46 & 11.07\\
atmosmodd & 4 & 7 & 7 & 6.94 & 0.24\\
pwtk & 2 & 180 & 54 & 53.39 & 4.74\\
\bottomrule
\end{tabular}
\end{minipage} 
\end{table*}

% \begin{table*}[t]
% \caption{Row-statistics of the matrices used in our evaluation.}
% \label{spconf:tab:matrix-stats}
% \begin{minipage}{.5\linewidth}
% \centering
% \footnotesize
% \begin{tabular}{lrrrrr}
% \toprule
% \textbf{Matrix} & \textbf{Min} & \textbf{Max} & \textbf{Med.} & \textbf{Mean} & \textbf{Std. Dev.}\\
% \midrule
% bcsstk17 & 1 & 150 & 45 & 39.06 & 15.41\\
% scircuit & 1 & 353 & 5 & 5.61 & 4.39\\
% mac\_econ\_fwd500 & 1 & 44 & 5 & 6.17 & 4.44\\
% majorbasis & 6 & 11 & 11 & 10.94 & 0.42\\
% Lin & 4 & 7 & 7 & 6.90 & 0.31\\
% rma10 & 4 & 145 & 46 & 50.69 & 27.78\\
% cop20k\_A & 0 & 81 & 23 & 21.65 & 13.79\\
% webbase-1M & 1 & 4700 & 2 & 3.11 & 25.35\\
% \bottomrule
% \end{tabular}
% \end{minipage}%
% \begin{minipage}{.5\linewidth}
% \centering
% \footnotesize
% \begin{tabular}{lrrrrr}
% \toprule
% \textbf{Matrix} & \textbf{Min} & \textbf{Max} & \textbf{Med.} & \textbf{Mean} & \textbf{Std. Dev.}\\
% \midrule
% cant & 1 & 78 & 74 & 64.17 & 14.06\\
% pdb1HYS & 18 & 204 & 123 & 119.31 & 31.86\\
% ecology1 & 3 & 5 & 5 & 5 & 0.06\\
% largebasis & 4 & 14 & 12 & 12.64 & 1.15\\
% consph & 1 & 81 & 81 & 72.13 & 19.08\\
% shipsec1 & 24 & 102 & 54 & 55.46 & 11.07\\
% atmosmodd & 4 & 7 & 7 & 6.94 & 0.24\\
% pwtk & 2 & 180 & 54 & 53.39 & 4.74\\
% \bottomrule
% \end{tabular}
% \end{minipage} 
% \end{table*}

The Figure~\ref{spconf:fig:matrices-histo} shows the histograms of the number of non-zero elements in each row of the matrices, and the Table~\ref{spconf:tab:matrix-stats} shows the row statistics of the matrices used in this evaluation. Evidently, \textit{ecology1} and \textit{largebasis} have low number of non-zeros per row, hence better search performance. Although \textit{shipsec1} has a higher number of non-zeros per row, these are clustered around the diagonal and have some structure in the matrix, making it easier for the binary search to break from the search early on. When the non-zeros are clustered around the diagonal and the number of non-zeros per row/column is relatively low, the search operation over the indexing array can quickly find the matching index or conclude that there is no matching index.

We see that there are a high number of non-zero values, considering both mean and median, per row in \textit{pdb1HYS}, resulting in higher search time. Hence, performing poorly compared to the best baseline. Although there are lower number of non-zeros per row in \textit{scircuit}, the non-zero elements are scattered along the row, making it difficult for the search to exit early on. When the non-zeros are scattered across the matrix or the number of non-zeros per row/column is high, the search operation, although logarithmic in complexity, takes more time to find the matching index or conclude that there is no matching index. Hence, the performance of \system is better when the non-zeros are clustered around the diagonal and the number of non-zeros per row/column is low. A performance engineer or an auto-scheduler~\cite{willowahrens,sparseauto} can use this insight to choose between \system and the baselines based on the sparsity patterns of the input matrices, or save some metadata with the matrices so that the right choice can be made automatically based on the statistics of the input matrices.

\begin{figure*}[t]
\begin{subfigure}{0.49\textwidth}
    \centering
    \includegraphics[width=\linewidth]{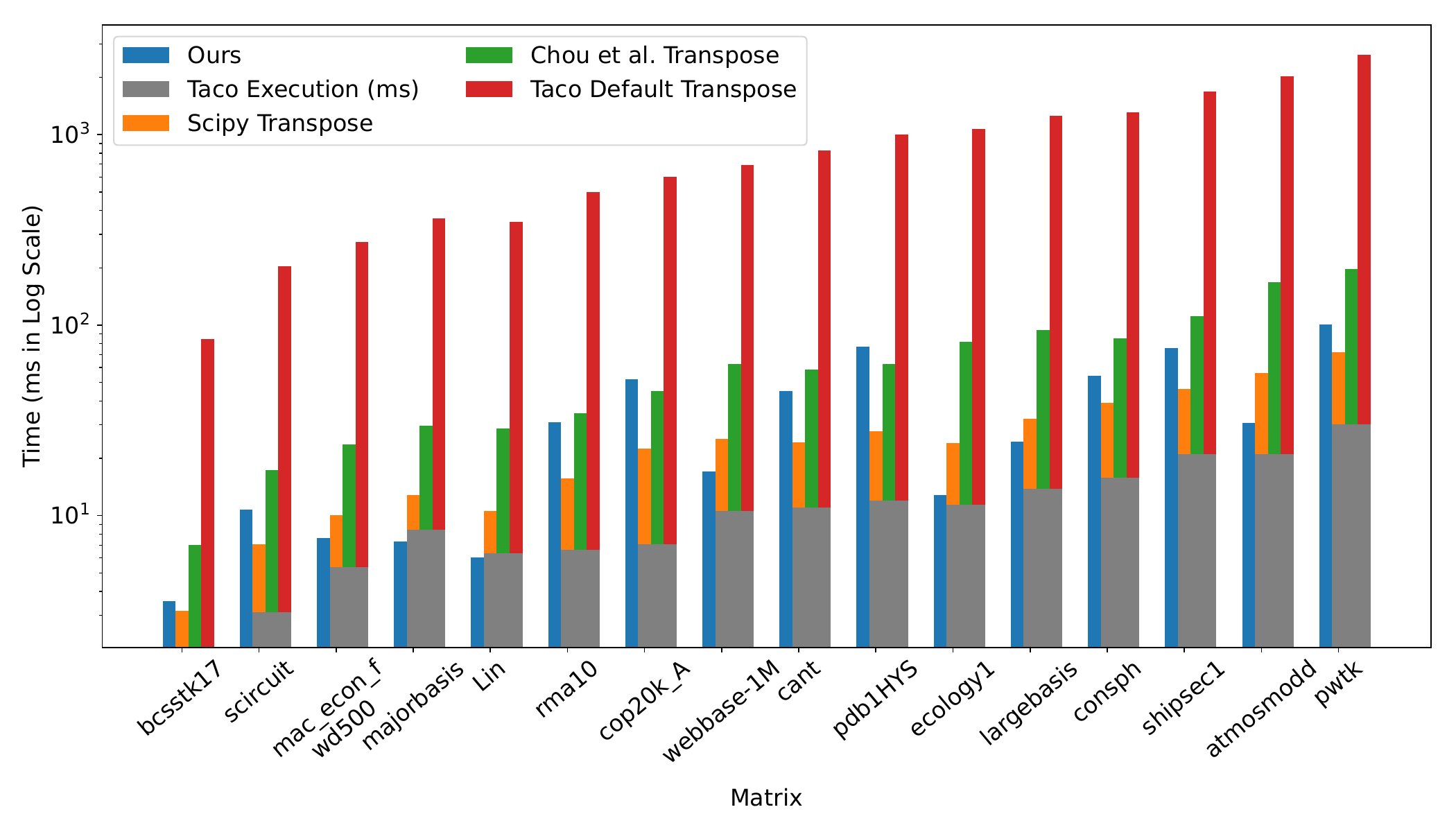}
    \caption{Execution time.}
    \label{spconf:fig:hadamard-contract-dense-exec-time}
\end{subfigure}%
\begin{subfigure}{0.49\textwidth}
    \centering
    \includegraphics[width=\linewidth]{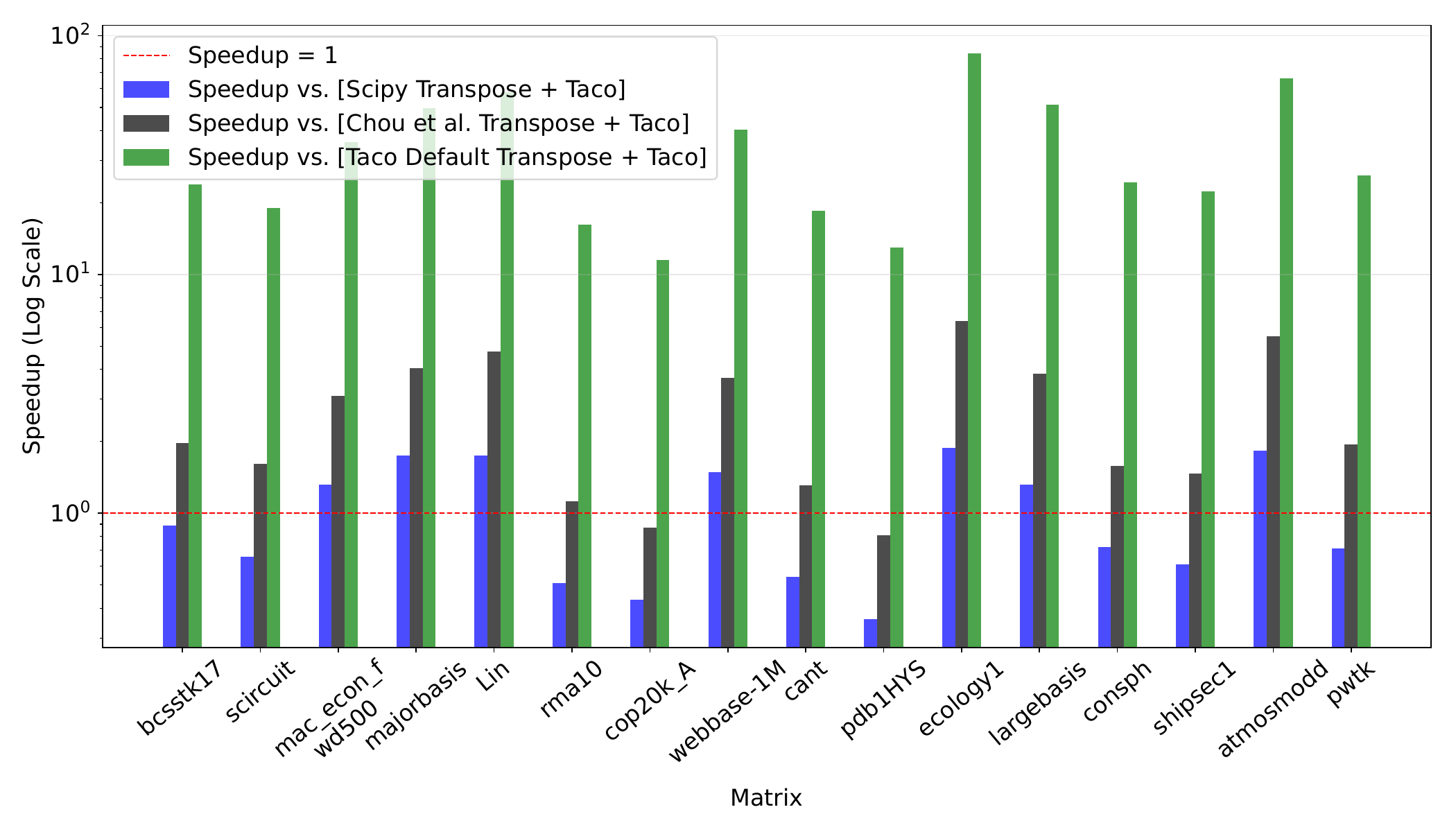}
    \caption{Speedup (Baseline Time / \system Time).}
    \label{spconf:fig:hadamard-contract-dense-speedup}
\end{subfigure}%
\caption{Performance comparison of a dot-product contraction kernel: $a(i) = {\sparse B(i,j)}*{\sparse C(j,i)}$. The output can be easily assembled in these kernels since it is dense in these examples, and we only report the time taken to compute the output values.}
\label{spconf:fig:hadamard-contract-dense}
\end{figure*}

\begin{figure*}[t]
\begin{subfigure}{0.49\textwidth}
    \centering
    \includegraphics[width=\linewidth]{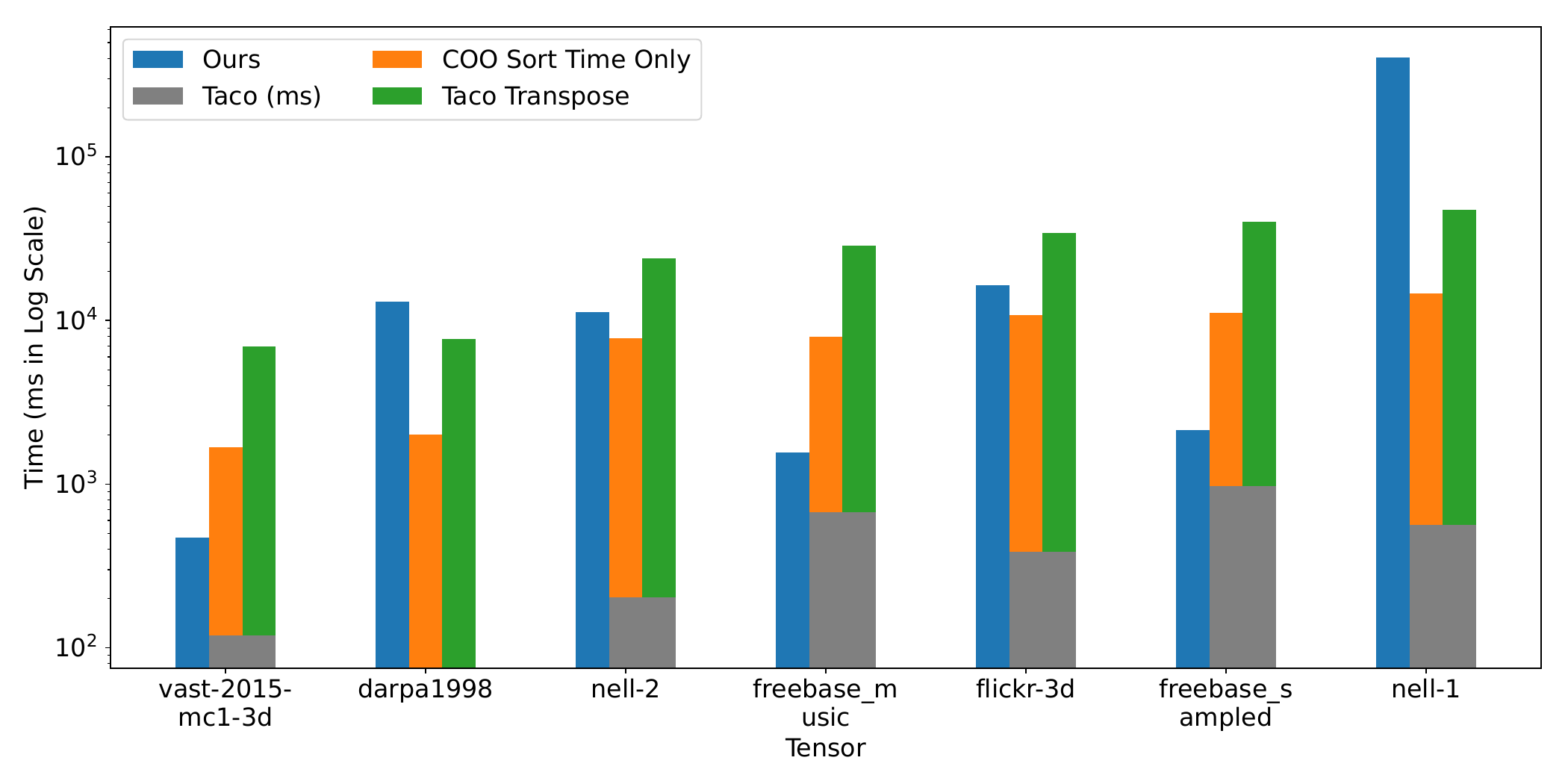}
    \caption{Execution time.}
    \label{spconf:fig:tensor-contract-1dout-exec-time}
\end{subfigure}%
\begin{subfigure}{0.49\textwidth}
    \centering
    \includegraphics[width=\linewidth]{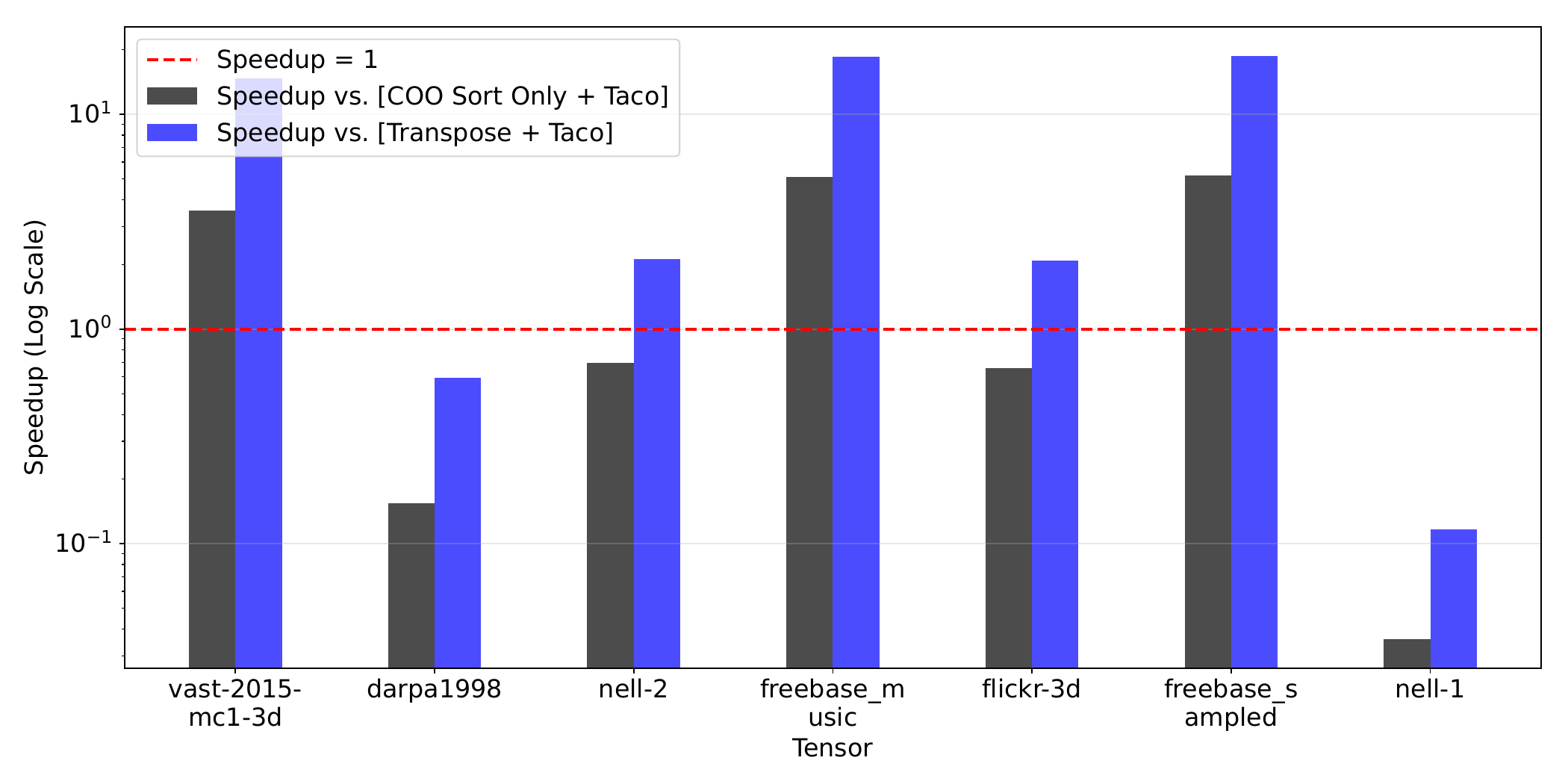}
    \caption{Speedup (Baseline Time / \system Time).}
    \label{spconf:fig:tensor-contract-1dout-speedup}
\end{subfigure}%
%\end{subcaptiongroup}
\caption{Performance comparison of a 3D tensor contraction kernel: $a(i) = {\sparse B(i,j,k)}*{\sparse C(i,k,j)}$.}
\label{spconf:fig:tensor-contract-1dout}
\end{figure*}

% corrected upto this point

Furthermore, we evaluate two other tensor contractions that require searching over one of the indexing arrays of the input tensors. The results are shown in Figures~\ref{spconf:fig:hadamard-contract-dense} and~\ref{spconf:fig:tensor-contract-1dout}. We use the same matrix/tensor for both inputs to ensure matching dimensions in the computation. The dot-product kernel results in Figure~\ref{spconf:fig:hadamard-contract-dense} show results similar to the elementwise multiplication kernel. Therefore, the results can be analyzed and explained using the same arguments as previously mentioned. 

The 3D tensor contraction kernel results in Figure~\ref{spconf:fig:tensor-contract-1dout} show mixed results across different datasets. We have two baselines in this evaluation. The first one uses TACO to transpose the tensor. The second one only benchmarks the time it takes to sort, using quick sort, the indices of an index and value tuples. For example if the coordinate values of the tensor are [(1,1,2,v1), (1,1,0,v2), (1,0,0,v3)] with indexing pattern (i,k,j) and we want to transpose it to have the (i,j,k) order, then the output after the sorting operation would be [(1,0,0,v3), (1,0,1,v2), (1,2,1,v1)]. Notice how the (j,k) position indices are changed in the output array. The performance of \system compared to the baselines depends on the sparsity patterns of the input tensors, and the dimensionality (bounds of the tensors, especially the tensor dimension which the search operation is performed) of the tensors. For example, in the case of \textit{vast-2015-mc1-3d}, the search is done over the third dimension which has a bound of 2, adding almost no overhead to the search operation. Hence, \system outperforms the baselines. The sparsities of \textit{freebase\_music} and \textit{freebase\_sampled} are extremely high, making the search operation take a shorter time. However, the search dimensionality of \textit{nell-1}, and \textit{nell-2} are higher compared to the other two dimensions, making the search operation take a longer time. Hence, \system performs poorly compared to the baselines in these two cases.

\begin{figure*}
\centering
\begin{subfigure}[c]{0.45\textwidth}
\begin{lstlisting}[tabsize=2, morekeywords={}]
// define tensors (not shown here)
// define the computation
A(i,j) = B(i,k) * C(k,j);
// define workspace intermediate temporary
TensorVar w("w",
    Type(A.getType().getDataType(), 
    {A.getType().getShape().getDimension(1)}),
    taco::dense);
// define the nested multi-child/inner loop structure in lower IR
IndexStmt stmt = forall(i,
    where(
        forall(j,
            A(i,j) = w(j)),
        forall(k,
            forall(j,
                w(j) += B(i,k) * C(k,j)))));
A.compile(stmt);
A.evaluate();
\end{lstlisting}
\caption{TACO SpMM using lower-level IR.}
\label{spconf:fig:taco-spmm-code}
\end{subfigure}
\begin{subfigure}[c]{0.45\textwidth}
\begin{lstlisting}[]
// define tensors (not shown here)
// define the computation
A(i,j) = B(i,k) * C(k,j);
IndexStmt stmt = A.getAssignment().concretize(sparseFused=true);
A.compile(stmt);
A.evaluate();
\end{lstlisting}
\caption{\system implementation.}
\label{spconf:fig:our-spmm-code}
\end{subfigure}
\caption{Code snippets for SpMM using TACO's lower level IR and \system.}
\label{spconf:fig:spmm-code-snippets}
\end{figure*}

\begin{figure*}[t]
\centering
\begin{subfigure}[c]{0.55\textwidth}
    \centering
    \includegraphics[width=\linewidth]{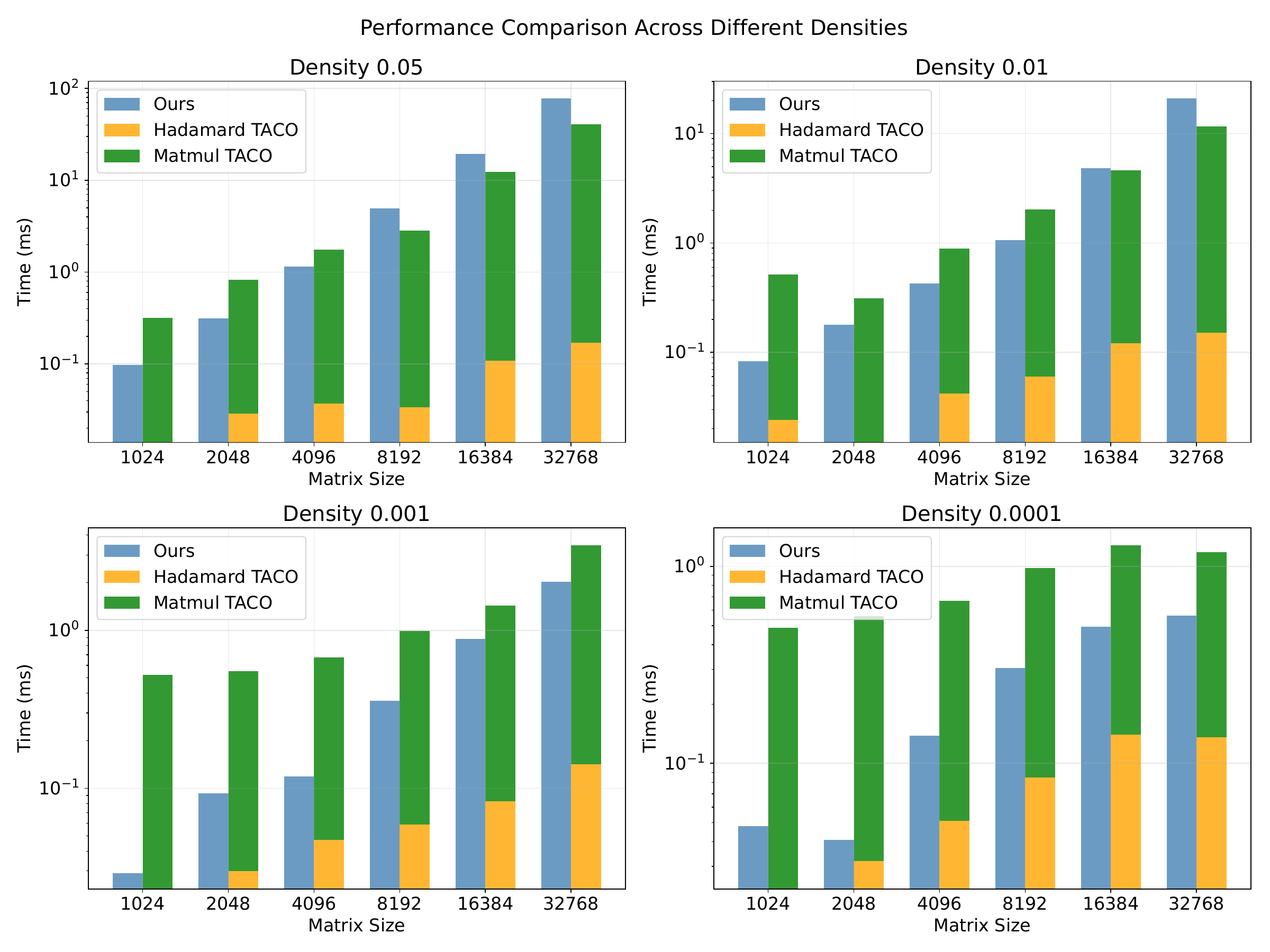}
    \caption{}
    \label{spconf:fig:hadamard-spmm-bar-plot}
\end{subfigure}%
\hspace{1em}%
\begin{subfigure}[c]{0.40\textwidth}
    \centering
    \includegraphics[width=1.1\linewidth]{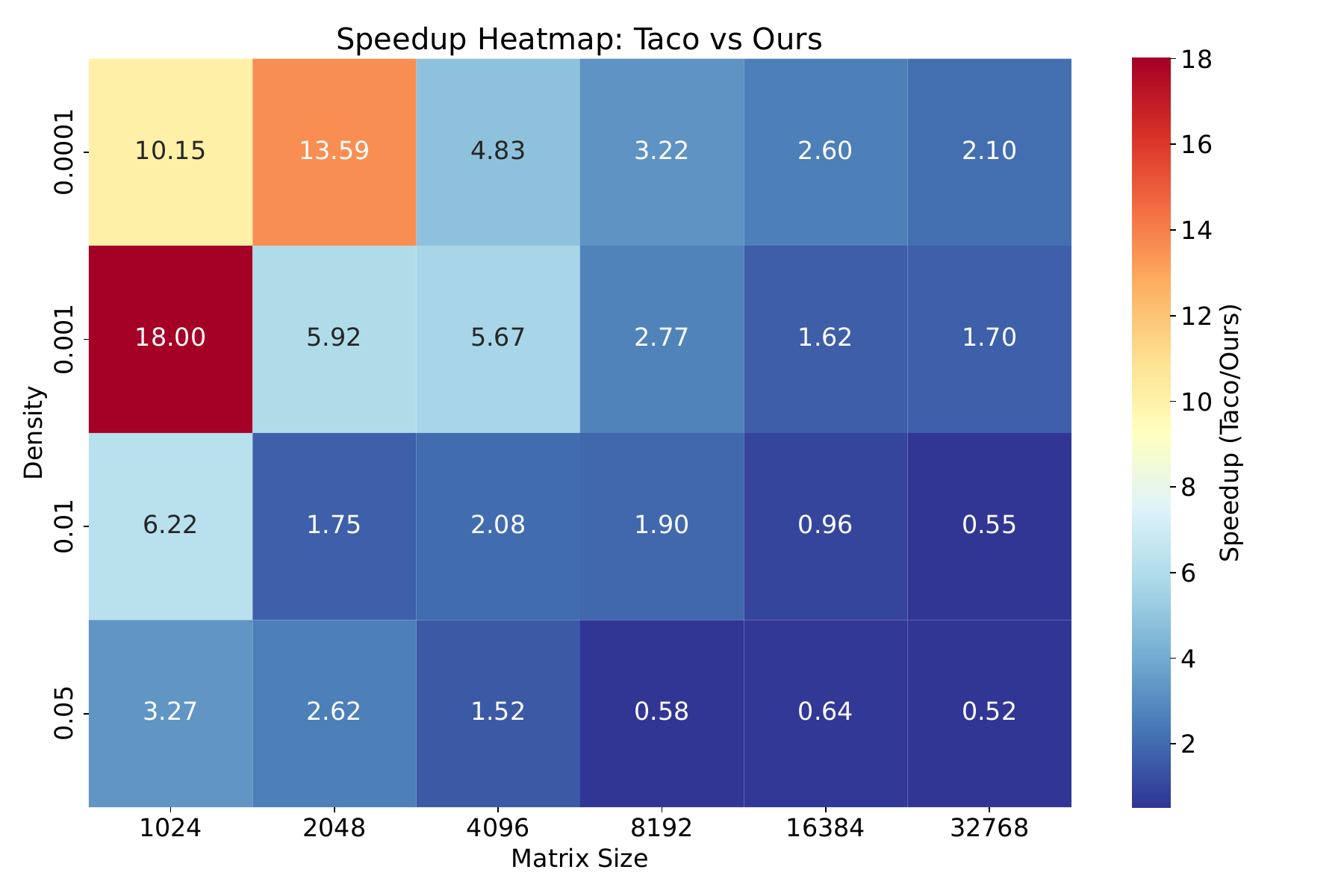}
    \caption{}
    \label{spconf:fig:hadamard-spmm-heatmap}
\end{subfigure}
\caption{Combination of elementwise multiplication and SpMM. ${\sparse A(i,j)} = {\sparse B(i,k)}*{\sparse C(k,j)}*{\sparse D(k,j)}$. Figure~\ref{spconf:fig:hadamard-spmm-bar-plot} shows the breakdown of time taken for elementwise multiplication and SpMM. Figure~\ref{spconf:fig:hadamard-spmm-heatmap} shows the speedup of our approach for different matrix sizes and densities.}
\label{spconf:fig:hadamard-spmm-plots}
\end{figure*}

\subsection{Case Study 2: Tensor contractions that require reordering to a branched loop nest}\label{spconf:case_study_2}

This section focuses on two cases where the tensor contraction requires reordering to a nested multi-child/inner loop structure. We evaluate the following kernels for this case study:
\begin{description}
    \item[SpMM] Sparse matrix-sparse matrix multiplication (SpMM) \\${\sparse A(i,j)} = {\sparse B(i,k)} * {\sparse C(k,j)}$
    \item[Elmul-SpMM] Combination of elementwise multiplication and sparse matrix-sparse matrix multiplication \\${\sparse A(i,j)} = {\sparse B(i,k)} * {\sparse C(k,j)} * {\sparse D(k,j)}$
\end{description}

TACO has a hardcoded pattern match for SpMM to rewrite the computation to lower level IR that represents a nested multi-child/inner loop structure. Therefore, we do not compare with TACO for SpMM. We qualitatively compare our implementation with TACO's current implementation in the discussion below. 

The pattern matching in TACO only works when the input tensors are in a specific order. In einsum notation, both ${\sparse A(i,j)} = {\sparse B(i,k)} * {\sparse C(k,j)}$ and ${\sparse A(i,j)} = {\sparse C(k,j)} * {\sparse B(i,k)}$ are semantically equivalent, and should yield the same result. However, TACO's pattern matching is brittle and only works for the first case. It generates incorrect code for the second case. If the user wanted to use the lower level IR that represents the nested multi-child/inner loop structure, it would require the user to write the code snippet as shown in Figure~\ref{spconf:fig:taco-spmm-code} where as \system only requires the user to write the code snippet as shown in Figure~\ref{spconf:fig:our-spmm-code}. Writing the lower level IR code snippet is tedious and requires the user to have a good understanding of TACO's lower level IR. Our implementation only requires a few lines of code (LoC), whereas the TACO lower level IR implementation requires comparatively more LoC.

The second example in this case study is a combination of elementwise multiplication and SpMM. We use the baseline below for this comparison.

\begin{description}
\item[TACO Split-Computation] In this setting, we split a larger tensor contraction
into two smaller tensor contractions to be used with default TACO compiler.
\end{description}

The computation we use here is ${\sparse A(i,j)} = {\sparse B(i,k)} * {\sparse C(k,j)} * {\sparse D(k,j)}$. We compare our implementation with a split computation that first computes the elementwise multiplication ${\sparse D(k,j)} = {\sparse C(k,j)} * {\sparse D(k,j)}$ and then computes the SpMM ${\sparse A(i,j)} = {\sparse B(i,k)} * {\sparse D(k,j)}$ using TACO. Figure~\ref{spconf:fig:hadamard-spmm-bar-plot} shows the breakdown of time taken for elementwise multiplication and SpMM. Figure~\ref{spconf:fig:hadamard-spmm-heatmap} shows the speedup of our approach for different matrix sizes and densities. Our approach outperforms the split computation in cases where the input matrices have high sparsities or when they are smaller in size.

\section{RELATED WORK}\label{spconf:related_works}

There has been significant prior work on optimizing sparse tensor algebra computations. Works such as TACO~\cite{taco, kjolstad:2018:workspaces, senanayake:2020:scheduling}, COMET~\cite{compiler_in_mlir}, Sparsifier~\cite{aart22} in MLIR, SparseTIR~\cite{ye2023sparsetir}, Galley~\cite{galley}, and Finch~\cite{finch2025} have contributed to various aspects of compiler optimizations in sparse tensor algebra. But none of these works have explored the idea of generating code without having to explicitly transpose tensors when there are conflicting iteration orders in the computation, and they have relied on transpose operations or temporary data structures such as hash maps to resolve these conflicts. 

Several prior works have explored optimizations for sparse tensor transpositions. Sparse Tensor Transpositions~\cite{mueller:2020:transposition} presents two parallel sorting algorithms to efficiently transpose sparse tensors stored in the coordinate format. Format Abstraction for Sparse Tensor Algebra Compilers~\cite{chou} introduces a level format abstraction to represent various compressed storage formats that can be easily composed with sparse tensor compilers. They also present an automatic code generation approach for sparse format conversion~\cite{chou2020-format-conversion}. UniSparse~\cite{unisparse2024} is another work that introduces custom data layouts for sparse tensors and provides support for tensor algebra code generation for these custom data layouts. These works primarily focus on customizing sparse data formats, providing ways for format conversion or transpositions, and optimizing the transposition operation itself, rather than exploring ways to avoid explicit transpositions in tensor computations. These works complement the previous works on sparse tensor compilers.
\section{DISCUSSION AND CONCLUSION}\label{spconf:conclusion}

Designing sparse tensor algebra compilers is a challenging problem due to the complexity of compressed storage formats and the non-affine loop nests they create. Since no single schedule is optimal for all sparsity patterns, it is crucial for compilers to express a wide range of schedules.

In this paper, we presented \system, a sparse tensor algebra compiler that addresses this by generating code for tensor contractions with conflicting iteration orders without explicitly transposing tensors. This is achieved by fusing the transposition into the main computation, which avoids materializing large temporary tensors and is particularly advantageous in memory-constrained environments. Our evaluation shows that for certain sparsity patterns, this fused approach significantly outperforms traditional transpose-and-compute strategies. However, this method introduces search overheads, creating a trade-off that depends on the tensor properties.

This work opens a clear path for future research, most notably the development of an auto-scheduling framework with a cost model to automatically choose the best strategy. In conclusion, \system expands the scheduling possibilities for sparse tensor algebra, enabling higher performance across a broader range of computations.

% \section{Data Availability}

% https://conf.researchr.org/track/splash-2026/oopsla-2026#data-availability-statement

% \ad{To help readers understand the state of the intended artifact, we ask you to add a section just before references titled Data-Availability Statement in the initial submission.}

% \ad{In it, indicate whether an artifact exists, its nature and limitations, and whether it will be submitted for Artifact Evaluation. This section should ideally also include links to preliminary versions of (anonymized) artifacts, datasets, and so on that reviewers may find useful (but are not obliged to follow).}

\section*{Data Availability Statement}
The paper will be accompanied by an artifact %~\cite{anon-artifact}
that contains the source code of \system, the benchmarks, and the scripts to reproduce the results presented in the paper.

\bibliography{references.bib}

\end{document}